\documentclass[12pt,aps,prl,onecolumn,showpacs,superscriptaddress,amsmath,amssymb]{revtex4-2}

\usepackage{setspace}

\usepackage{graphicx}
\usepackage{dcolumn}
\usepackage{bm}
\usepackage{amsmath}
\usepackage{amssymb}
\usepackage{graphicx}
\usepackage{indentfirst}
\usepackage{booktabs}
\usepackage{multirow}
\usepackage{colortbl}
\usepackage{epsfig}

\begin{document}

\onehalfspacing

\title{High-order AMR in two-dimensional magnetic monolayers from spin mixing}

\author{M. Q. Dong}
\affiliation{State Key Laboratory for Mechanical Behavior of Materials, School of Materials Science and Engineering, Xi’an Jiaotong University, Xi’an, Shaanxi, 710049, China.}
\affiliation{Key Laboratory of Computational Physical Sciences (Ministry of Education), Institute of Computational Physical Sciences, State Key Laboratory of Surface Physics, and Department of Physics, Fudan University, Shanghai 200433, China.}


\author{Zhi-Xin Guo}
\email{zxguo08@xjtu.edu.cn}
\affiliation{State Key Laboratory for Mechanical Behavior of Materials, School of Materials Science and Engineering, Xi’an Jiaotong University, Xi’an, Shaanxi, 710049, China.}

\date{\today}

\begin{abstract}

Anisotropic magnetoresistance (AMR) is a well-known magnetoelectric coupling phenomenon, commonly exhibiting two-fold symmetry relative to the magnetic field. In this study, we reveal the existence of high-order AMRs in two-dimensional (2D) magnetic monolayers. Based on density functional theory (DFT) calculations of Fe$_3$GeTe$_2$ and CrTe$_2$ monolayers, we find that different energy bands contribute uniquely to AMR behavior. The high-order AMR is attributed to strong spin mixing at band crossing points, which induces significant Berry curvature. This curvature also contributes to the AMR for electrons with dominant spin-up or spin-down polarization characteristics. However, for electrons exhibiting strong spin mixing, the Berry curvature effect becomes nontrivial, resulting in high-order AMR. Our findings provide an effective approach to identifying and optimizing materials with high-order AMR, which is critical for designing high-performance spintronic devices.  

\end{abstract}

\maketitle



Magnetoresistances (MRs) encompass a wide range of phenomena, and anisotropic magnetoresistance (AMR) has been extensively studied \cite{mcguire1975anisotropic, doring1938abhangigkeit, I_A_Campbell_1970, ritzinger2023anisotropic, PhysRevLett.125.097201, WilliamThomson1857}. AMR refers to the response of charge carriers in magnetic materials to changes in magnetization direction. The traditional explanation of AMR in ferromagnetic metals and alloys attributes it to spin-orbit interaction (mixed spin-up and spin-down states) and the anisotropic scattering probabilities of charge carriers \cite{mcguire1975anisotropic, doring1938abhangigkeit}. Phenomenologically, the angular dependence of resistivity ($\rho$) is commonly expressed as:
\begin{eqnarray}
	\frac{\rho _{xx}}{\rho _0}=1+C_I\cos2\alpha,
	\label{eq:amr}
\end{eqnarray}
where $C_I$ is sometimes called non-crystalline AMR since it persists in polycrystalline systems, and $\alpha$ is the angle between the current and magnetization \cite{ritzinger2023anisotropic}. This equation, which shows a two-fold symmetry nature of AMR with respect to  $\alpha$, generally applies to polycrystalline and amorphous samples. On the other hand, D\"oring's  studied \cite{doring1938abhangigkeit, doring1960richtungsabhangigkeit} on AMR in Fe and Ni single crystals further explored AMR as a function of $\alpha$ and the angle of magnetization (\textbf{M}) relative to crystallographic directions.  In addition to the non-crystalline AMR, the crystalline AMR containing high-order (containing $\cos4\alpha$ and $\cos6\alpha$ terms) was observed, which was generally  attributed to the macroscopic crystal symmetry \cite{doring1938abhangigkeit, doring1960richtungsabhangigkeit}.  However, a theoretical proof is still exile, especially from the microscopic viewpoint.

During the past decades, AMR has been widely reported in conventional transition metals and their alloys \cite{doring1938abhangigkeit, WilliamThomson1857, PhysRev.165.670, tomlinson1882iv, PhysRevB.45.9819, PhysRevB.58.6434}, dilute magnetic semiconductors \cite{PhysRevB.65.212407, PhysRevB.72.085201, jungwirth2002boltzmann}, film of ferromagnetic metals and semiconductors \cite{ma2023anisotropic, sun2022anisotropic, sun2020room}, with values typically around a few percent. Recently, two-dimensional (2D)  magnetic monolayers have become a focal point in condensed matter physics research due to their intriguing properties in spintronics \cite{apl_lb, li2022intriguing, PhysRevB.109.205105, PhysRevB.103.094433, PhysRevB.102.115413}. Despite the fact that AMR in 2D magnetic monolayers has been rarely studied, these materials actually provide an ideal platform for exploring intriguing magnetic properties \cite{burch2018magnetism}, making them particularly suitable for studying the microscopic theory behind the high-order AMR effect. 

In this work, we combine DFT and Boltzmann transport equation (BTE) methods to investigate AMR in monolayer Fe$_3$GeTe$_2$ (FGT) and CrTe$_2$. We observe that AMR behavior varies with chemical potential (Fermi energies), and high-order AMR occurs at chemical potential where band crossing and spin state mixing. We find that the large gradients in Berry curvature near band crossing points can induce complex electron behavior, leading to the observed high-order AMR.


We performed DFT calculations on monolayer Fe$_3$GeTe$_2$  and CrTe$_2$ using the projector augmented wave (PAW) method \cite{PhysRevB.59.1758, PhysRevB.50.17953}, implemented in the Vienna ab initio simulation package (VASP) \cite{PhysRevB.54.11169, PhysRevB.47.558}. Following the DFT calculations, maximally localized Wannier functions (MLWFs) \cite{PhysRev.52.191, RevModPhys.84.1419, PhysRevB.56.12847} were constructed using the Wannier90 code \cite{MOSTOFI2008685, Pizzi_2020}. The electrical conductivity was subsequently calculated using the BTE method \cite{Ziman_1972, grosso2013solid, PIZZI2014422}, with the chemical potential ($\mu$) and temperature ($T$) dependence of electrical conductivity described as:
\begin{eqnarray}
\sigma _{ij}\left( \mu ,T \right) =e^2\int_{-\infty}^{+\infty}{d\varepsilon \left( -\frac{\partial f\left( \varepsilon ,\mu ,T \right)}{\partial \varepsilon} \right) \Sigma_{ij}\left(\varepsilon\right)},
	\label{eq:sigma}
\end{eqnarray}
where $f\left( \varepsilon ,\mu ,T \right)$ is the Fermi-Dirac distribution function,
\begin{eqnarray}
	f\left( \varepsilon ,\mu ,T \right)=\frac{1}{e^{\left(\varepsilon-\mu\right)/k_{\bm{B}}T}+1},
	\label{eq:fd}
\end{eqnarray}
and $\Sigma_{ij}\left(\varepsilon\right)$ is the transport distribution function (TDF) tensor, defined as
\begin{eqnarray}
	\Sigma_{ij}\left(\varepsilon\right)=\frac{1}{V}\sum_{n,\bf{k}}{v_i\left(n,\bf{k}\right)v_j\left(n,\mathbf{k}\right)\tau\left(n,\bf{k}\right)\delta\left(\varepsilon-E_{n,\bf{k}}\right)}.
	\label{eq:tdf}
\end{eqnarray}
The sum is taken over all energy bands ($n$) and states ($\mathbf{k}$), see Supplementary Material for details \cite{supplementary}.


Monolayer Fe$_3$GeTe$_2$ (FGT) is known to be phase-stable and can maintain long-range ferromagnetic (FM) order at reduced temperatures around 130 K, compared to approximately 200 K in bulk FGT \cite{deng2018gate, fei2018two,
PhysRevB.93.134407, deiseroth2006fe3gete2}. Figures 1(a) and 1(b) illustrate the atomic structure of monolayer FGT, showing three Fe atoms per unit cell located in two inequivalent Wyckoff positions. The calculated lattice constants for monolayer FGT yield a=b=3.91 \AA. Furthermore, the magnetic moments of the Fe atoms were found to be 1.73 $\mu_B$ and 1.01 $\mu_B$ for the two inequivalent sites, respectively, which is in good agreement with previous studies \cite{PhysRevB.93.134407}.

\begin{figure}
	\centering
	\includegraphics[width=\linewidth]{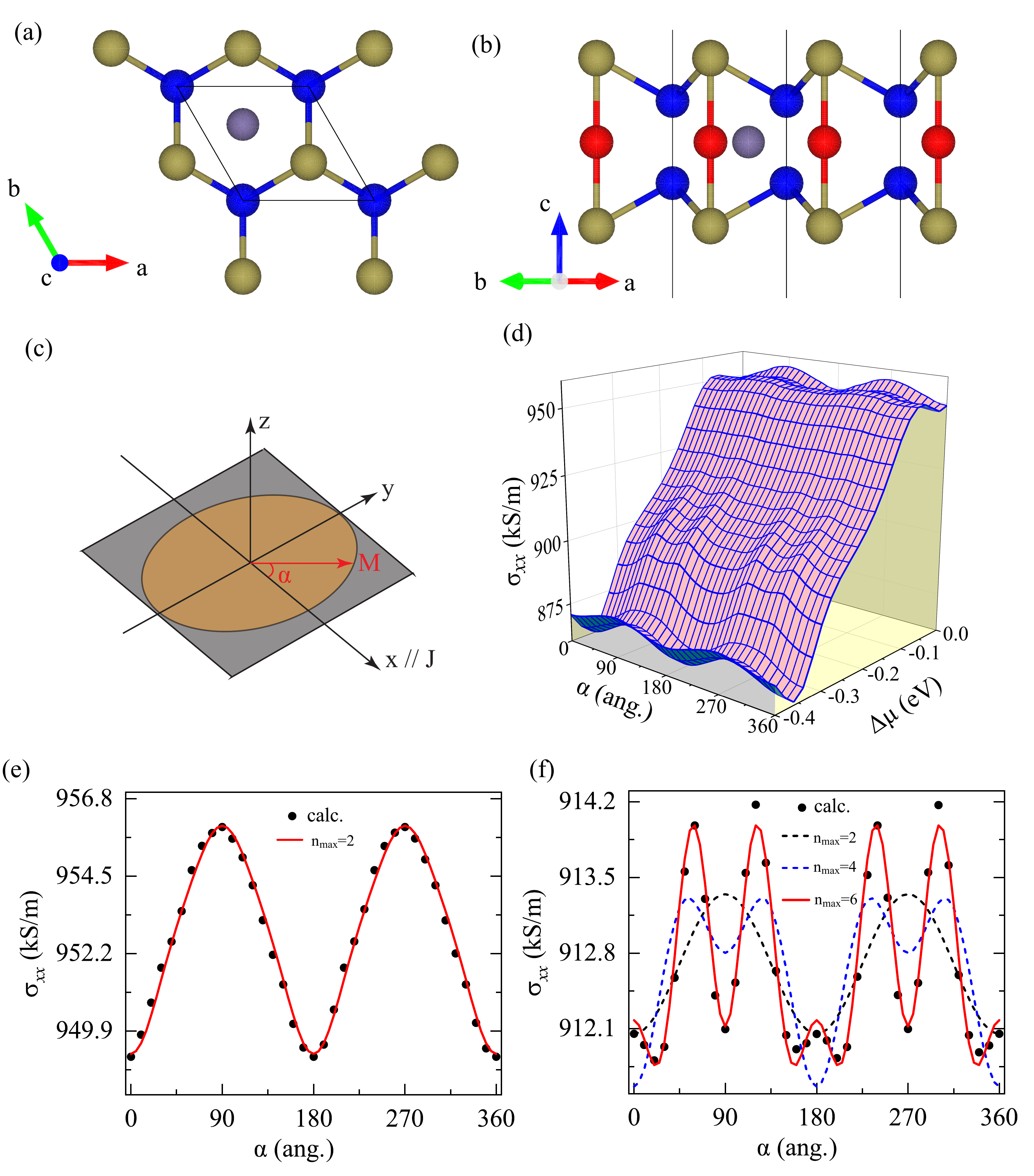}
	\caption{\label{fig:fig1}(a)-(b) Top and side views of atomic structures of monolayer Fe$_3$GeTe$_2$, where red and blue balls represent two distinct Fe atoms, silver balls represents Ge atoms, and light green balls represents Te atoms. (c) Schematic diagram defining the angle $\alpha$ in Cartesian coordinates. (d) Electronic conductivity ($\sigma_{xx}$) for different magnetization directions and chemical potentials $\Delta\mu$. (e)–(f) $\sigma_{xx}$ for different magnetization directions at $\Delta\mu$ = 0 eV (e) and $\Delta\mu$ = – 0.2 eV (f), respectively. In (e), expanding the AMR up to $\cos(2\alpha)$ is sufficient to replicate the computed results, while, in (f) a good fitting can be obtained only when the expanding is up to $\cos(6\alpha)$.}
\end{figure}

The in-plane AMR, where the magnetization rotates in the $x-y$ plane ($\alpha$, Fig. 1(c)), is the primary focus of our investigation. The $\alpha$-dependent electrical conductivity ($\sigma_{xx}$) was computed for various chemical potentials ($\Delta\mu=\mu-E_F$). As shown in Fig. 1(d), the magnitude of AMR is highly dependent on the chemical potential, with significant AMR observed in the energy ranges of [-0.4, -0.2] eV and [-0.1, 0] eV, while negligible AMR appears in the range of [-0.2, -0.1] eV. This behavior aligns with our previous studies on CrPX$_3$ monolayers and CoFe bulk systems, where band splitting related to chemical potential explained the AMR variation \cite{PhysRevB.108.L020401, hou2024giantanisotropicmagnetoresistancemagnetic}.

Interestingly, the number of peaks in the $\sigma_{xx}(\alpha)$ curves is also strongly dependent on the chemical potential ($\Delta\mu$). For example, at $\Delta\mu$ within the range [-0.4, -0.3] eV and [-0.1, 0] eV, two peaks are observed, showing a functional form of $\sigma_{xx}=A\,\cos2\alpha$, consistent with non-crystalline AMR behavior. However, at $\Delta\mu$ within the range [-0.3, -0.15] eV, four peaks emerge, suggesting high-order AMR characteristic. To illustrate this phenomenon more clearly, Figs. 1(e) and 1(f) show the AMR at two specific chemical potentials: $\Delta\mu$ = 0 eV (non-crystalline AMR region) and $\Delta\mu$ = -0.2 eV (high-order AMR region). The former exhibits a two-fold symmetry, while the latter reveals additional high-order components.

To further understand these results, we conducted a Fourier component analysis, decomposing the computed data into a sum of cosine terms \cite{gonzalez2024anisotropic}:
\begin{eqnarray}
	\sigma_{xx}\left(\alpha\right)\approx\sum_nC_n\cos\left(n\alpha\right),
	\label{eq:fourier}
\end{eqnarray}
where $C_n$ represents the amplitude of the $n$-th harmonic. As shown in Fig. 1(e), for $\Delta\mu$ = 0 eV, expanding the AMR up to $\cos(2\alpha)$ is sufficient to replicate the computed results, confirming a two-fold symmetry. However, for $\Delta\mu$ = -0.2 eV (Fig. 1(f)), a $\cos(2\alpha)$ expansion alone can not capture the data accurately. Instead, a $\cos(6\alpha)$ term is necessary to achieve a good fit, indicating the presence of high-order symmetry. Table 1 presents the fitted $C_n$ values for these two chemical potentials, showing that at $\Delta\mu$ = -0.2 eV, the $C_2$, $C_4$, and $C_6$ harmonics are of comparable magnitude, reflecting significant four-fold and six-fold symmetry in the AMR.
\begin{table}[b]
	\caption{\label{tab:table1}
		Amplitudes $C_n$ for different harmonics.
	}
	\begin{ruledtabular}
		\begin{tabular}{r*{4}{c}}
			$\Delta\mu$&$C_0$&$C_2$&$C_4$&$C_6$ \\
			\colrule 
			-0.2 & 952.8 & -3.236 & -0.1915 & -0.1567 \\
			0 & 912.7 & -0.6519 & -0.5489 & 0.6811 \\		
		\end{tabular}
	\end{ruledtabular}
\end{table}

We next investigated the microscopic mechanism underlying the high-order AMR observed in monolayer FGT. According to semi-classical wave packet theory, electrons move on an isoenergetic surface that is perpendicular to the applied magnetic field. Figure 2 shows the calculated shapes of the isoenergetic surface at $\Delta\mu$ = 0 eV and $\Delta\mu$ = -0.2 eV for different magnetization rotation angles $\alpha$. Unexpectedly, the isoenergetic surfaces exhibit six-fold symmetry for both chemical potentials, regardless of angle $\alpha$. This suggests that the shape of  isoenergetic surface does not directly correspond to the AMR in magnetic materials, contrasting with earlier studies on nonmagnetic materials in a magnetic field \cite{PhysRevB.99.035142}.

\begin{figure}
	\centering
	\includegraphics[width=\linewidth]{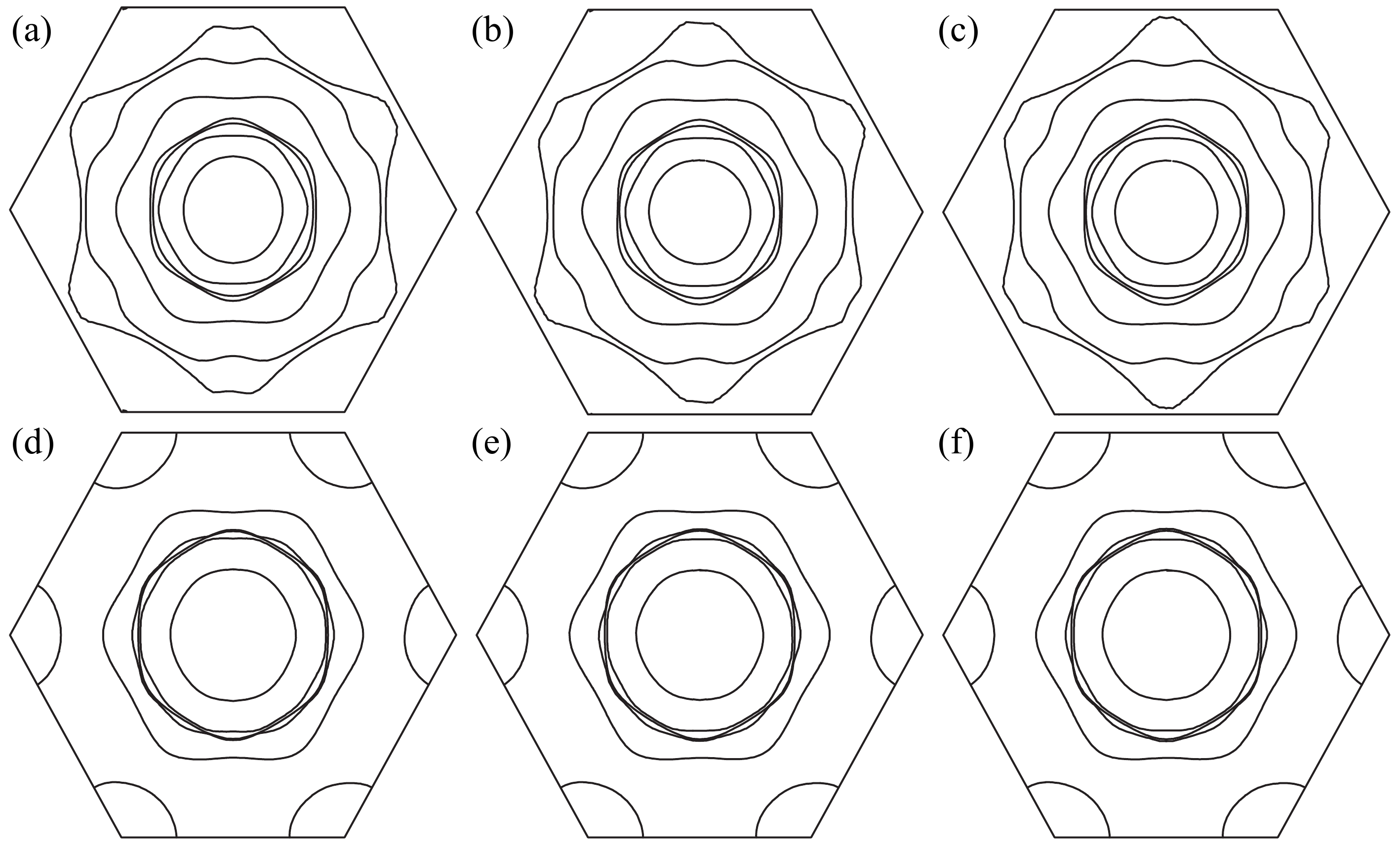}
	\caption{\label{fig:fig2}Magnetization direction dependent Fermi surface of monolayer FGT at  $\Delta\mu$ = 0 eV (a)-(c) and $\Delta\mu$ = -0.2 eV (d)-(f). (a) and (d) for $\alpha$ = 0$^\circ$, (b) and (e) for $\alpha$ = 60$^\circ$, (c) and (f) for $\alpha$=90$^\circ$.}
\end{figure}

According to the BTE Eqs. \ref{eq:amr}-\ref{eq:fd}, electrical conductance directly correlates with electron velocity, which is the gradient of the energy band ($\partial E/\hbar\partial k$). Consequently, it is not the shapes of the isoenergetic surface, but rather the changes in the isoenergetic surface with respect to $\alpha$ that should be more directly correlated with variations in electrical conductance and, consequently, AMR. Our numerical calculations further confirm this. As seen in Fig. 2, the variation in the isoenergetic surface with respect to $\alpha$ is more significant at $\Delta\mu$ = 0 eV than at $\Delta\mu$ = -0.2 eV, suggesting that the AMR is more pronounced at $\Delta\mu$= 0 eV. This is supported by the much greater variation in $\sigma_{xx}$ with respect to $\alpha$ at $\Delta\mu$= 0 eV than at $\Delta\mu$= -0.2 eV (Fig. 1).

\begin{figure}
	\centering
	\includegraphics[width=\linewidth]{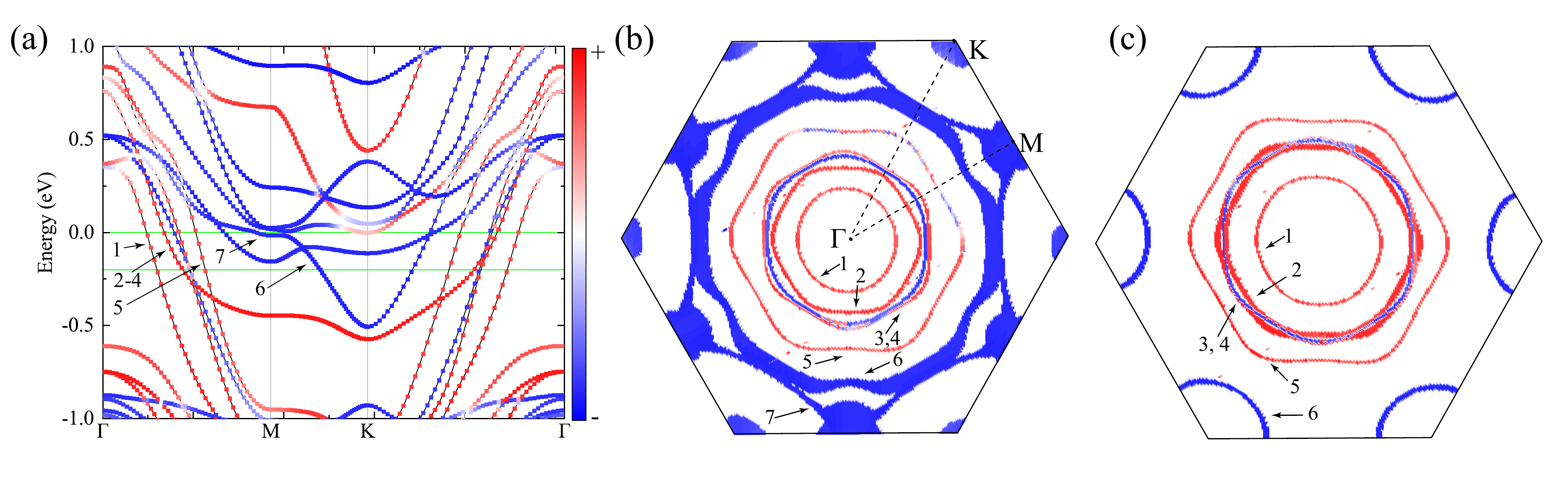}
	\caption{\label{fig:fig3}Band structure of spin projection for monolayer FGT (a), and spin projections on the First Brillouin Zone isoenergetic plane at $\Delta\mu$= 0 eV (b) and $\Delta\mu$= - 0.2 eV (c), respectively. In (a), seven special bands are marked by index 1-7, two green lines mark the chemical potentials of $\Delta\mu$= 0 eV and $\Delta\mu$= - 0.2 eV, respectively. In (b) and (c), the red and blue lies represent spin-up and spin-down projections, respectively.}
\end{figure}

Next, we explored the relationship between band structure and high-order AMR. Since electrical conductance at a given chemical potential can be considered the sum of contributions from all nearby electronic states, we analyzed the contribution of individual energy bands to the AMR:
\begin{eqnarray}
	\sigma_{xx}\left(\mu\right)=\sum_n\sigma_{xx}^n\left(\mu\right),
	\label{eq:sigmaxx}
\end{eqnarray}
where $\sigma_{xx}^n$ represents the conductivity contributed by $n$-th band, and $n$ is the index of energy band. Figure 3(a) shows that seven energy bands cross the isoenergetic surface at $\Delta\mu$= 0 eV, whereas six bands cross at $\Delta\mu$= -0.2 eV. We further plot the variation of electrical conductance with $\alpha$ for each of these bands in Fig. 4. At $\Delta\mu$= 0 eV, all bands contribute to a two-fold symmetry in conductance, consistent with the overall AMR. In contrast, at $\Delta\mu$= -0.2 eV, while three bands (indexes 1, 2, 5 in Fig. 3(a)) maintain two-fold symmetry, two bands (indexes 3, 4 in Fig. 3(a)) exhibit a more complex behavior, deviating from two-fold symmetry and driving the high-order AMR. Notably, one band (index 6 in Fig. 3(a)) contributes minimally to the AMR at $\Delta\mu$= -0.2 eV, owing to its negligible effect on electrical conductance. This feature indicates that the high-order AMR is induced by some energy bands with certain characteristics.
\begin{figure}
	\centering
	\includegraphics[width= 0.5\linewidth]{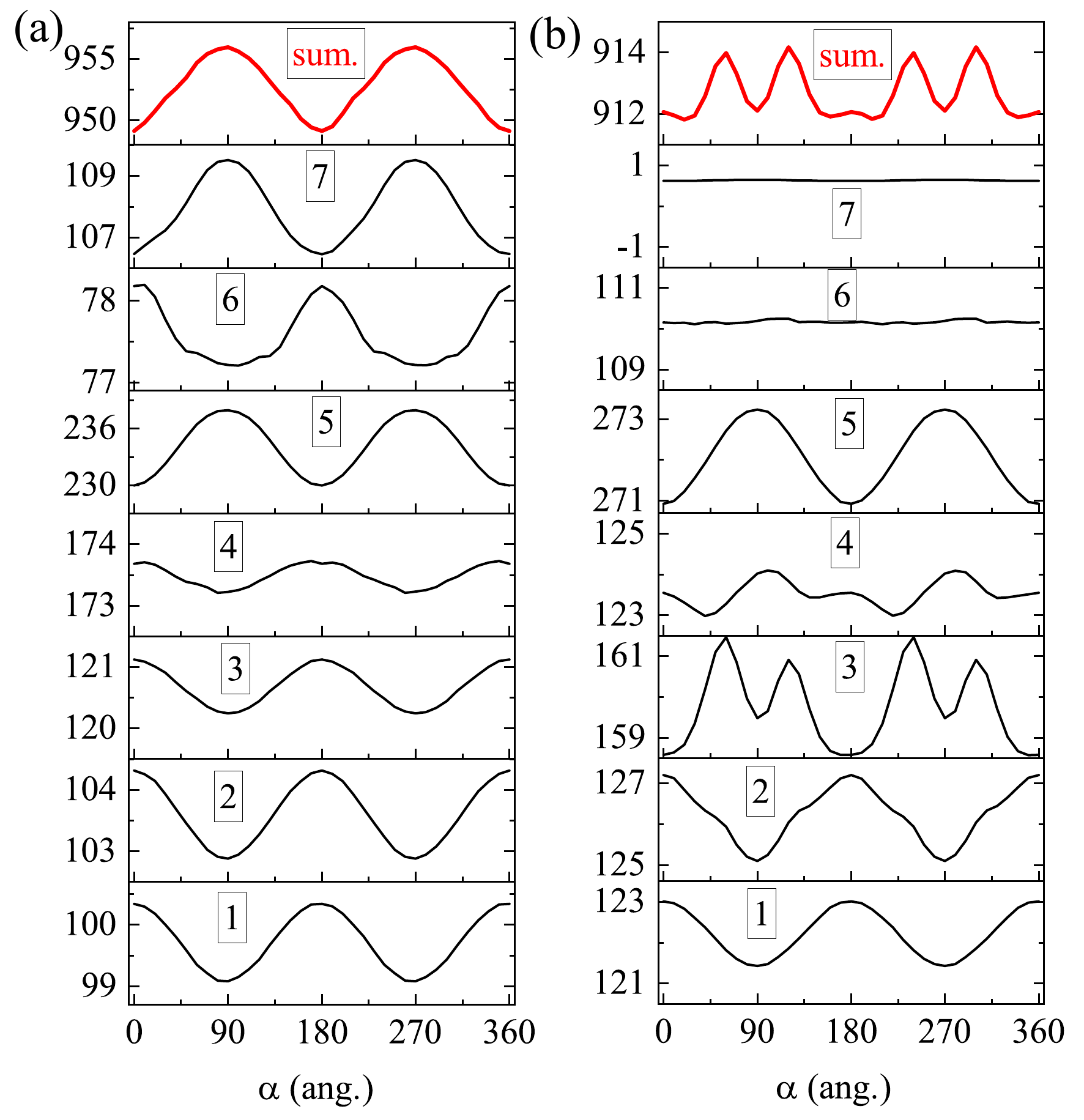}
	\caption{\label{fig:fig4} The electrical conductivity at two different chemical potentials $\Delta\mu$ for monolayer FGT. The red solid line represents the total conductivity, and the black lines represent the conductivity from each band with the same index marked in Fig. 3(a). (a) In case of  $\Delta\mu$= 0 eV, and (b) In case of $\Delta\mu$= -0.2 eV.}
\end{figure}
To further analyze the origin of high-order AMR, we examined the two energy bands contributing to this behavior, i.e., indexes 3 and 4 as indicated in Fig. 3 and Fig. 4(b). 

We observed that the two energy bands responsible for high-order AMR at $\Delta\mu$= -0.2 eV do not exhibit the same behavior at $\Delta\mu$ = 0 eV (Fig. 4(a)). This difference arises from the spin-polarized band dispersions, which vary with position in the First Brillouin Zone (FBZ). At $\Delta\mu$= 0 eV, the two spin-polarized bands along the $\Gamma-\text{M}$ high symmetric path are well separated, whereas at $\Delta\mu$= -0.2 eV, they exhibit significant overlap due to dispersion effects.

To validate this argument, we calculated spin-projected isoenergetic contours in the FBZ for both $\Delta\mu$= 0 eV and $\Delta\mu$=-0.2 eV. As shown in Fig. 3(b), at $\Delta\mu$=0 eV, the spin-polarized isoenergetic contours of the selected energy bands (indexes 3 and 4) do not overlap, while at $\Delta\mu$= -0.2 eV, these contours show significant overlap. This overlap results in noticeable mixing between spin-up states and spin-down states (referred to as spin mixing), particularly in the crossover regions (see Fig. 3(c)). Such spin mixing is further corroborated by the calculated Kohn–Sham (KS) orbitals \cite{li2017topological} of the two relevant bands. Figure 5 illustrates that at $\Delta\mu$=0 eV, the energy band of index 3 (4) exhibit nearly pure spin-down (spin-up) state, whereas at $\Delta\mu$=-0.2 eV, it additionally displays sizable spin-up (spin-down) spin states, respectively, at the crossover points along the $\Gamma-\text{M}$ direction.

\begin{figure}
	\centering
	\includegraphics[width=\linewidth]{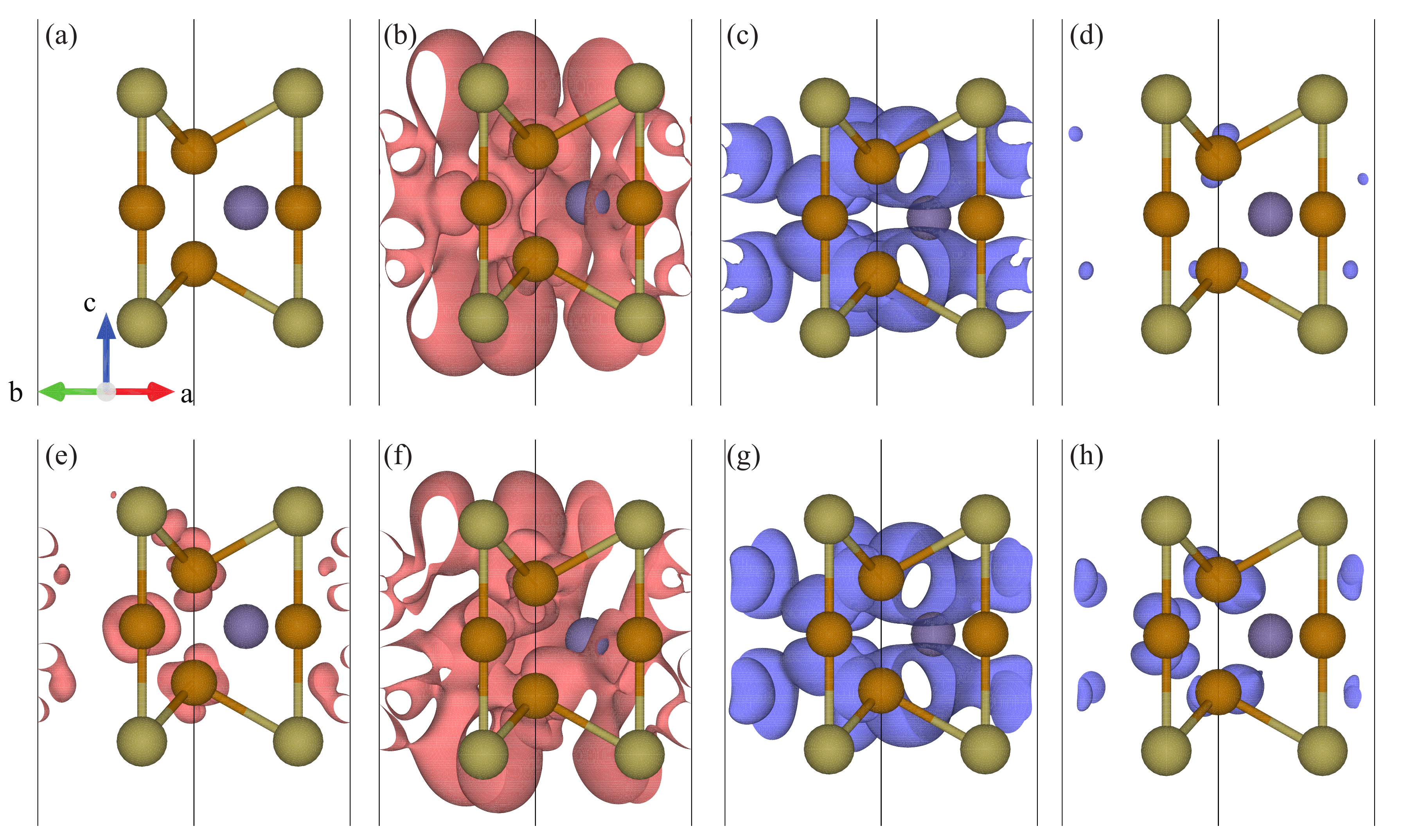}
	\caption{\label{fig:fig5}Calculated Kohn–Sham (KS) orbitals at $\Delta\mu$=0 eV (a - d) and $\Delta\mu$= -0.2 eV (e - h) for energy bands of index 3 (a, c, e, g) and index 4 (b, d, f, h) of monolayer FGT, with an isosurface value of $3\times 10^{-10}$. Red and blue colors indicate spin-up and spin-down projections, respectively. At $\Delta\mu$=0 eV, energy band of index 3 exhibits only spin-down characteristic (a, c), while at $\Delta\mu$= -0.2 eV, it additionally presents some  spin-up characteristic (e, g). Similarly, at $\Delta\mu$=0 eV, energy band of index 4 exhibits nearly pure spin-up characteristic (b, d), whereas at $\Delta\mu$= -0.2 eV the spin-down characteristic becomes sizable (f, h).}
\end{figure}

The above analysis suggests that the combination of band crossover and spin mixing at specific chemical potentials is key to the emergence of high-order AMR. However, it remains unclear whether these two effects contribute independently or if their interplay is essential. To address this question, we further investigated AMR in scenarios where only one of these effects is present: (1) band crossover without spin mixing and (2) spin mixing without band crossover. For this analysis, we selected monolayer CrTe$_2$ (Fig. S2) , a simpler system with a more straightforward band structure compared to FGT, allowing us to isolate each condition.

\begin{figure}
	\centering
	\includegraphics[width=\linewidth]{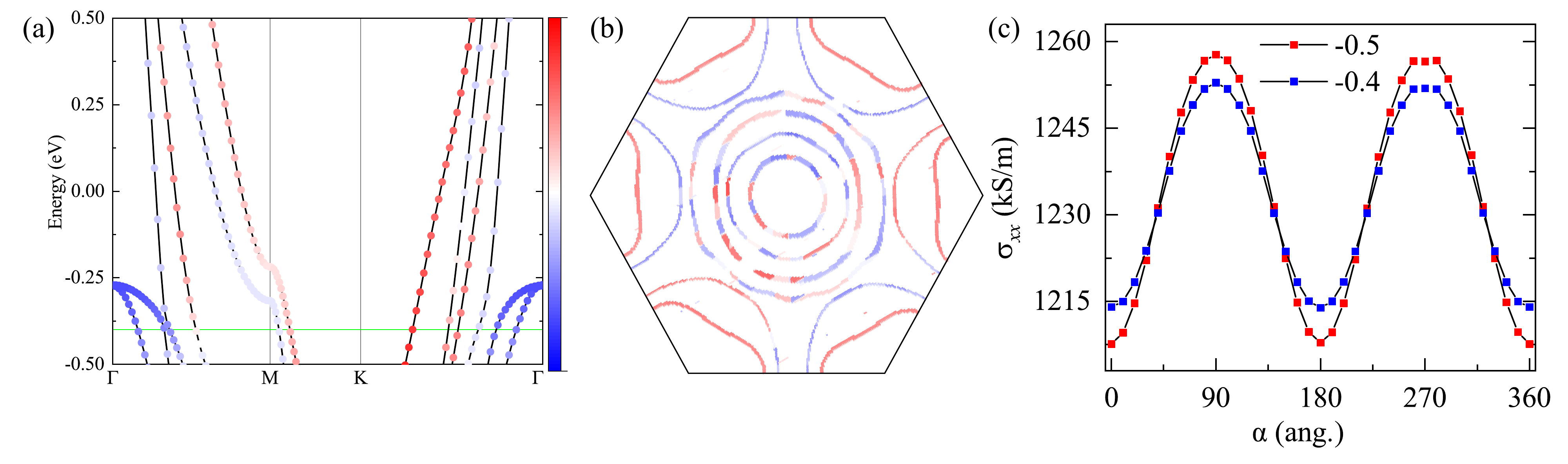}
	\caption{\label{fig:fig6}Band structure of spin projection for monolayer CrTe$_2$ (a), spin projections on the First Brillouin Zone isoenergetic plane at $\Delta\mu$= - 0.5 eV (b), $\sigma_{xx}$ for different magnetization directions at $\Delta\mu$ = -0.4 eV and $\Delta\mu$ = – 0.5 eV. Green line in (a) indicates the chemical potential at $\Delta\mu$= -0.4 eV.}
\end{figure}

Figure 6(a) first shows the case of band crossover without spin mixing at $\Delta\mu$=-0.4 eV. The AMR observed here remains two-fold symmetric, as shown in Fig. 6(c), indicating that band crossover alone is insufficient to induce high-order AMR. Conversely, Fig. 6(b) presents a scenario with strong spin mixing at $\Delta\mu$=-0.5 eV in the two-dimensional Brillouin Zone without band crossover. Again, the calculated AMR retains a two-fold symmetry, confirming that spin mixing by itself does not produce high-order AMR. These results clearly demonstrate that the emergence of high-order AMR requires a synergistic interaction between band crossover and spin mixing, both occurring at the same energy level and momentum, which significantly influence the magnetization-dependent electronic structure.

Finally, we provide a phenomenological explanation for the microscopic mechanism of high-order AMR by incorporating Berry curvature corrections into the group velocity \cite{PhysRevB.59.14915, PhysRevB.53.7010, PhysRevLett.92.037204}. The equations of motion, modified to include Berry curvature $\Omega_n(\textbf{k})$, are given by:

\begin{align}
	\dot{\mathbf{x}}=\frac{1}{\hbar}\frac{\partial
		\varepsilon_n(\mathbf{k})}{\partial\mathbf{k}}
		-\dot{\mathbf{k}}\times
		\mathbf{\Omega}_n(\mathbf{k}), \notag \\
	\hbar\dot{\mathbf{k}}=-e\mathbf{E}-e\dot{\mathbf{x}},
	\times \mathbf{B},
	\label{eq:berrycurv}
\end{align}

where $\mathbf{\Omega}_n(\mathbf{k})$ represents the Berry curvature of the Bloch state. This curvature functions analogously to a magnetic field in momentum space \cite{PhysRevB.59.14915}, much like the real-space magnetic field $\mathbf{B}$. In real space, a nonzero gradient of $\mathbf{B}$ causes spin-up and spin-down  electrons to acquire opposite acceleration \cite{gerlach1989experimentelle, RevModPhys.87.1213}. Similarly, a gradient in  $\mathbf{\Omega}_n(\mathbf{k})$($\nabla\mathbf{\Omega}_n(\mathbf{k})$) can impart differing acceleration to electrons with distinct spins.

In the case of monolayer FGT, only the $\mathbf{\Omega}_n^z(\mathbf{k})$ component of the Berry curvature is nonzero, which can be computed by standard methods \cite{PhysRevLett.92.037204}:

\begin{align}
	\mathbf{\Omega}_n^z(\mathbf{k}) &=-\sum_{n^\prime\ne n}
	\frac{2\text{Im}\langle \psi_{n\mathbf{k}}|v_x|
		\psi_{n^\prime \mathbf{k}}\rangle
	\langle \psi_{n^\prime \mathbf{k}}|v_y|\psi_{n\mathbf{k}}
	\rangle}{\left(\omega_{n^\prime}-\omega_n\right)^2},\notag \\
	\mathbf{\Omega}^z(\mathbf{k}) &=\sum_n 
	\mathbf{\Omega}_n^z(\mathbf{k})
	\label{eq:bcnz}
\end{align}

Due to the inverse relationship between $\left(\omega_{n^\prime}-\omega_n\right)$ and $\mathbf{\Omega}_n^z(\mathbf{k})$, significant values of $\mathbf{\Omega}_n^z(\mathbf{k})$ and its gradient $\nabla\mathbf{\Omega}_n^z(\mathbf{k})$ are observed near band crossing points (see Fig. S2). This large gradient gives rise to a significant difference in the acceleration of spin-up and spin-down electrons, leading to a corresponding change in their velocity. This change affects the electric current in the $x$-direction and, thus, contributes to the AMR.

As shown in Fig. S2, the magnitude of $\mathbf{\Omega}^z(\mathbf{k})$ and $\nabla\mathbf{\Omega}^z(\mathbf{k})$ near band crossing points is notably sensitive to the magnetization direction (\textbf{M}), highlighting the influence of Berry curvature on AMR. For electrons with a predominant spin-up or spin-down polarization, changes in velocity are synchronized with variations in $\nabla\mathbf{\Omega}^z(\mathbf{k})$ and, consequently, the magnetization direction. However, for electrons with significant spin mixing (i.e., a mixture of spin-up state and spin-down states), the response of velocity to changes in $\nabla\mathbf{\Omega}^z(\mathbf{k})$ is more complex. This interaction results in the emergence of high-order AMR. 


In summary, our combined DFT and BTE calculations have revealed the presence of high-order AMR in 2D magnetic monolayers. We have shown that different energy bands contribute distinctively to the AMR behavior, with high-order AMR primarily arising from the interplay between band crossover and spin mixing at specific chemical potentials. Neither band crossover nor spin mixing alone is sufficient to induce high-order AMR; their synergistic interaction is crucial. Our analysis suggests that large gradients in Berry curvature near band crossing points may induce complex electron behavior, leading to the observed high-order AMR. These findings offer a promising pathway for identifying and optimizing materials with enhanced AMR properties, providing crucial insights for the development of high-performance spintronic devices.


\begin{acknowledgments}
We are grateful for valuable discussions with Professor Xiangrong Wang, and acknowledge financial support from the Natural Science Foundation of China (Grant No. 12474237, 12074301), and Science Fund for Distinguished Young Scholars of Shaanxi Province (No. 2024JC-JCQN-09). 
\end{acknowledgments}


\begin{thebibliography}{50}%
	\makeatletter
	\providecommand \@ifxundefined [1]{%
		\@ifx{#1\undefined}
	}%
	\providecommand \@ifnum [1]{%
		\ifnum #1\expandafter \@firstoftwo
		\else \expandafter \@secondoftwo
		\fi
	}%
	\providecommand \@ifx [1]{%
		\ifx #1\expandafter \@firstoftwo
		\else \expandafter \@secondoftwo
		\fi
	}%
	\providecommand \natexlab [1]{#1}%
	\providecommand \enquote  [1]{``#1''}%
	\providecommand \bibnamefont  [1]{#1}%
	\providecommand \bibfnamefont [1]{#1}%
	\providecommand \citenamefont [1]{#1}%
	\providecommand \href@noop [0]{\@secondoftwo}%
	\providecommand \href [0]{\begingroup \@sanitize@url \@href}%
	\providecommand \@href[1]{\@@startlink{#1}\@@href}%
	\providecommand \@@href[1]{\endgroup#1\@@endlink}%
	\providecommand \@sanitize@url [0]{\catcode `\\12\catcode `\$12\catcode
		`\&12\catcode `\#12\catcode `\^12\catcode `\_12\catcode `\%12\relax}%
	\providecommand \@@startlink[1]{}%
	\providecommand \@@endlink[0]{}%
	\providecommand \url  [0]{\begingroup\@sanitize@url \@url }%
	\providecommand \@url [1]{\endgroup\@href {#1}{\urlprefix }}%
	\providecommand \urlprefix  [0]{URL }%
	\providecommand \Eprint [0]{\href }%
	\providecommand \doibase [0]{https://doi.org/}%
	\providecommand \selectlanguage [0]{\@gobble}%
	\providecommand \bibinfo  [0]{\@secondoftwo}%
	\providecommand \bibfield  [0]{\@secondoftwo}%
	\providecommand \translation [1]{[#1]}%
	\providecommand \BibitemOpen [0]{}%
	\providecommand \bibitemStop [0]{}%
	\providecommand \bibitemNoStop [0]{.\EOS\space}%
	\providecommand \EOS [0]{\spacefactor3000\relax}%
	\providecommand \BibitemShut  [1]{\csname bibitem#1\endcsname}%
	\let\auto@bib@innerbib\@empty
	\bibitem [{\citenamefont {McGuire}\ and\ \citenamefont
		{Potter}(1975)}]{mcguire1975anisotropic}%
	\BibitemOpen
	\bibfield  {author} {\bibinfo {author} {\bibfnamefont {T.}~\bibnamefont
			{McGuire}}\ and\ \bibinfo {author} {\bibfnamefont {R.}~\bibnamefont
			{Potter}},\ }\bibfield  {title} {\bibinfo {title} {Anisotropic
			magnetoresistance in ferromagnetic 3d alloys},\ }\href@noop {} {\bibfield
		{journal} {\bibinfo  {journal} {IEEE Transactions on Magnetics}\ }\textbf
		{\bibinfo {volume} {11}},\ \bibinfo {pages} {1018} (\bibinfo {year}
		{1975})}\BibitemShut {NoStop}%
	\bibitem [{\citenamefont {D{\"o}ring}(1938)}]{doring1938abhangigkeit}%
	\BibitemOpen
	\bibfield  {author} {\bibinfo {author} {\bibfnamefont {W.}~\bibnamefont
			{D{\"o}ring}},\ }\bibfield  {title} {\bibinfo {title} {Die abh{\"a}ngigkeit
			des widerstandes von nickelkristallen von der richtung der spontanen
			magnetisierung},\ }\href@noop {} {\bibfield  {journal} {\bibinfo  {journal}
			{Annalen der Physik}\ }\textbf {\bibinfo {volume} {424}},\ \bibinfo {pages}
		{259} (\bibinfo {year} {1938})}\BibitemShut {NoStop}%
	\bibitem [{\citenamefont {Campbell}\ \emph {et~al.}(1970)\citenamefont
		{Campbell}, \citenamefont {Fert},\ and\ \citenamefont
		{Jaoul}}]{I_A_Campbell_1970}%
	\BibitemOpen
	\bibfield  {author} {\bibinfo {author} {\bibfnamefont {I.~A.}\ \bibnamefont
			{Campbell}}, \bibinfo {author} {\bibfnamefont {A.}~\bibnamefont {Fert}},\
		and\ \bibinfo {author} {\bibfnamefont {O.}~\bibnamefont {Jaoul}},\ }\bibfield
	{title} {\bibinfo {title} {The spontaneous resistivity anisotropy in
			{Ni}-based alloys},\ }\href@noop {} {\bibfield  {journal} {\bibinfo
			{journal} {Journal of Physics C: Solid State Physics}\ }\textbf {\bibinfo
			{volume} {3}},\ \bibinfo {pages} {S95} (\bibinfo {year} {1970})}\BibitemShut
	{NoStop}%
	\bibitem [{\citenamefont {Ritzinger}\ and\ \citenamefont
		{V{\`y}born{\`y}}(2023)}]{ritzinger2023anisotropic}%
	\BibitemOpen
	\bibfield  {author} {\bibinfo {author} {\bibfnamefont {P.}~\bibnamefont
			{Ritzinger}}\ and\ \bibinfo {author} {\bibfnamefont {K.}~\bibnamefont
			{V{\`y}born{\`y}}},\ }\bibfield  {title} {\bibinfo {title} {Anisotropic
			magnetoresistance: materials, models and applications},\ }\href@noop {}
	{\bibfield  {journal} {\bibinfo  {journal} {Royal Society Open Science}\
		}\textbf {\bibinfo {volume} {10}},\ \bibinfo {pages} {230564} (\bibinfo
		{year} {2023})}\BibitemShut {NoStop}%
	\bibitem [{\citenamefont {Zeng}\ \emph {et~al.}(2020)\citenamefont {Zeng},
		\citenamefont {Ren}, \citenamefont {Li}, \citenamefont {Zeng}, \citenamefont
		{Jia}, \citenamefont {Miao}, \citenamefont {Hoffmann}, \citenamefont {Zhang},
		\citenamefont {Wu},\ and\ \citenamefont {Yuan}}]{PhysRevLett.125.097201}%
	\BibitemOpen
	\bibfield  {author} {\bibinfo {author} {\bibfnamefont {F.~L.}\ \bibnamefont
			{Zeng}}, \bibinfo {author} {\bibfnamefont {Z.~Y.}\ \bibnamefont {Ren}},
		\bibinfo {author} {\bibfnamefont {Y.}~\bibnamefont {Li}}, \bibinfo {author}
		{\bibfnamefont {J.~Y.}\ \bibnamefont {Zeng}}, \bibinfo {author}
		{\bibfnamefont {M.~W.}\ \bibnamefont {Jia}}, \bibinfo {author} {\bibfnamefont
			{J.}~\bibnamefont {Miao}}, \bibinfo {author} {\bibfnamefont {A.}~\bibnamefont
			{Hoffmann}}, \bibinfo {author} {\bibfnamefont {W.}~\bibnamefont {Zhang}},
		\bibinfo {author} {\bibfnamefont {Y.~Z.}\ \bibnamefont {Wu}},\ and\ \bibinfo
		{author} {\bibfnamefont {Z.}~\bibnamefont {Yuan}},\ }\bibfield  {title}
	{\bibinfo {title} {Intrinsic mechanism for anisotropic magnetoresistance and
			experimental confirmation in
			{${\mathrm{Co}}_{x}{\mathrm{Fe}}_{1\ensuremath{-}x}$} single-crystal films},\
	}\href@noop {} {\bibfield  {journal} {\bibinfo  {journal} {Physical Review
				Letters}\ }\textbf {\bibinfo {volume} {125}},\ \bibinfo {pages} {097201}
		(\bibinfo {year} {2020})}\BibitemShut {NoStop}%
	\bibitem [{\citenamefont {Thomson}(1857)}]{WilliamThomson1857}%
	\BibitemOpen
	\bibfield  {author} {\bibinfo {author} {\bibfnamefont {W.}~\bibnamefont
			{Thomson}},\ }\bibfield  {title} {\bibinfo {title} {Xix. on the
			electro-dynamic qualities of metals:—effects of magnetization on the
			electric conductivity of nickel and of iron},\ }\href@noop {} {\bibfield
		{journal} {\bibinfo  {journal} {Proceedings of the Royal Society of London}\
		}\textbf {\bibinfo {volume} {8}},\ \bibinfo {pages} {546} (\bibinfo {year}
		{1857})}\BibitemShut {NoStop}%
	\bibitem [{\citenamefont {D{\"o}ring}\ and\ \citenamefont
		{Simon}(1960)}]{doring1960richtungsabhangigkeit}%
	\BibitemOpen
	\bibfield  {author} {\bibinfo {author} {\bibfnamefont {W.}~\bibnamefont
			{D{\"o}ring}}\ and\ \bibinfo {author} {\bibfnamefont {G.}~\bibnamefont
			{Simon}},\ }\bibfield  {title} {\bibinfo {title} {Die
			richtungsabh{\"a}ngigkeit der magnetostriktion},\ }\href@noop {} {\bibfield
		{journal} {\bibinfo  {journal} {Annalen der Physik}\ }\textbf {\bibinfo
			{volume} {460}},\ \bibinfo {pages} {373} (\bibinfo {year}
		{1960})}\BibitemShut {NoStop}%
	\bibitem [{\citenamefont {Berger}\ and\ \citenamefont
		{Friedberg}(1968)}]{PhysRev.165.670}%
	\BibitemOpen
	\bibfield  {author} {\bibinfo {author} {\bibfnamefont {L.}~\bibnamefont
			{Berger}}\ and\ \bibinfo {author} {\bibfnamefont {S.~A.}\ \bibnamefont
			{Friedberg}},\ }\bibfield  {title} {\bibinfo {title} {Magnetoresistance of a
			permalloy single crystal and effect of $3d$ orbital degeneracies},\
	}\href@noop {} {\bibfield  {journal} {\bibinfo  {journal} {Physical Review}\
		}\textbf {\bibinfo {volume} {165}},\ \bibinfo {pages} {670} (\bibinfo {year}
		{1968})}\BibitemShut {NoStop}%
	\bibitem [{\citenamefont {Tomlinson}(1882)}]{tomlinson1882iv}%
	\BibitemOpen
	\bibfield  {author} {\bibinfo {author} {\bibfnamefont {H.}~\bibnamefont
			{Tomlinson}},\ }\bibfield  {title} {\bibinfo {title} {Iv. the influence of
			stress and strain on the action of physical forces},\ }\href@noop {}
	{\bibfield  {journal} {\bibinfo  {journal} {Proceedings of the Royal Society
				of London}\ }\textbf {\bibinfo {volume} {33}},\ \bibinfo {pages} {276}
		(\bibinfo {year} {1882})}\BibitemShut {NoStop}%
	\bibitem [{\citenamefont {Jen}(1992)}]{PhysRevB.45.9819}%
	\BibitemOpen
	\bibfield  {author} {\bibinfo {author} {\bibfnamefont {S.~U.}\ \bibnamefont
			{Jen}},\ }\bibfield  {title} {\bibinfo {title} {Anisotropic magnetoresistance
			of co-pd alloys},\ }\href@noop {} {\bibfield  {journal} {\bibinfo  {journal}
			{Physical Review B}\ }\textbf {\bibinfo {volume} {45}},\ \bibinfo {pages}
		{9819} (\bibinfo {year} {1992})}\BibitemShut {NoStop}%
	\bibitem [{\citenamefont {Chakraborty}\ and\ \citenamefont
		{Majumdar}(1998)}]{PhysRevB.58.6434}%
	\BibitemOpen
	\bibfield  {author} {\bibinfo {author} {\bibfnamefont {S.}~\bibnamefont
			{Chakraborty}}\ and\ \bibinfo {author} {\bibfnamefont {A.~K.}\ \bibnamefont
			{Majumdar}},\ }\bibfield  {title} {\bibinfo {title} {Galvanomagnetic studies
			in
			$\ensuremath{\gamma}\ensuremath{-}\mathrm{Ni}_{100\ensuremath{-}x\ensuremath{-}y}\mathrm{Fe}_{x}\mathrm{Cr}_{y}$
			permalloys $(5<~x<23; 2<~y<21)$},\ }\href@noop {} {\bibfield  {journal}
		{\bibinfo  {journal} {Physical Review B}\ }\textbf {\bibinfo {volume} {58}},\
		\bibinfo {pages} {6434} (\bibinfo {year} {1998})}\BibitemShut {NoStop}%
	\bibitem [{\citenamefont {Baxter}\ \emph {et~al.}(2002)\citenamefont {Baxter},
		\citenamefont {Ruzmetov}, \citenamefont {Scherschligt}, \citenamefont
		{Sasaki}, \citenamefont {Liu}, \citenamefont {Furdyna},\ and\ \citenamefont
		{Mielke}}]{PhysRevB.65.212407}%
	\BibitemOpen
	\bibfield  {author} {\bibinfo {author} {\bibfnamefont {D.~V.}\ \bibnamefont
			{Baxter}}, \bibinfo {author} {\bibfnamefont {D.}~\bibnamefont {Ruzmetov}},
		\bibinfo {author} {\bibfnamefont {J.}~\bibnamefont {Scherschligt}}, \bibinfo
		{author} {\bibfnamefont {Y.}~\bibnamefont {Sasaki}}, \bibinfo {author}
		{\bibfnamefont {X.}~\bibnamefont {Liu}}, \bibinfo {author} {\bibfnamefont
			{J.~K.}\ \bibnamefont {Furdyna}},\ and\ \bibinfo {author} {\bibfnamefont
			{C.~H.}\ \bibnamefont {Mielke}},\ }\bibfield  {title} {\bibinfo {title}
		{Anisotropic magnetoresistance in
			$\mathrm{Ga}_{1\ensuremath{-}x}\mathrm{Mn}_{x}\mathrm{As}$},\ }\href@noop {}
	{\bibfield  {journal} {\bibinfo  {journal} {Physical Review B}\ }\textbf
		{\bibinfo {volume} {65}},\ \bibinfo {pages} {212407} (\bibinfo {year}
		{2002})}\BibitemShut {NoStop}%
	\bibitem [{\citenamefont {Wang}\ \emph {et~al.}(2005)\citenamefont {Wang},
		\citenamefont {Edmonds}, \citenamefont {Campion}, \citenamefont {Zhao},
		\citenamefont {Foxon},\ and\ \citenamefont {Gallagher}}]{PhysRevB.72.085201}%
	\BibitemOpen
	\bibfield  {author} {\bibinfo {author} {\bibfnamefont {K.~Y.}\ \bibnamefont
			{Wang}}, \bibinfo {author} {\bibfnamefont {K.~W.}\ \bibnamefont {Edmonds}},
		\bibinfo {author} {\bibfnamefont {R.~P.}\ \bibnamefont {Campion}}, \bibinfo
		{author} {\bibfnamefont {L.~X.}\ \bibnamefont {Zhao}}, \bibinfo {author}
		{\bibfnamefont {C.~T.}\ \bibnamefont {Foxon}},\ and\ \bibinfo {author}
		{\bibfnamefont {B.~L.}\ \bibnamefont {Gallagher}},\ }\bibfield  {title}
	{\bibinfo {title} {Anisotropic magnetoresistance and magnetic anisotropy in
			high-quality $(\mathrm{Ga,Mn})\mathrm{As}$ films},\ }\href@noop {} {\bibfield
		{journal} {\bibinfo  {journal} {Physical Review B}\ }\textbf {\bibinfo
			{volume} {72}},\ \bibinfo {pages} {085201} (\bibinfo {year}
		{2005})}\BibitemShut {NoStop}%
	\bibitem [{\citenamefont {Jungwirth}\ \emph {et~al.}(2002)\citenamefont
		{Jungwirth}, \citenamefont {Abolfath}, \citenamefont {Sinova}, \citenamefont
		{Ku{\v{c}}era},\ and\ \citenamefont {MacDonald}}]{jungwirth2002boltzmann}%
	\BibitemOpen
	\bibfield  {author} {\bibinfo {author} {\bibfnamefont {T.}~\bibnamefont
			{Jungwirth}}, \bibinfo {author} {\bibfnamefont {M.}~\bibnamefont {Abolfath}},
		\bibinfo {author} {\bibfnamefont {J.}~\bibnamefont {Sinova}}, \bibinfo
		{author} {\bibfnamefont {J.}~\bibnamefont {Ku{\v{c}}era}},\ and\ \bibinfo
		{author} {\bibfnamefont {A.}~\bibnamefont {MacDonald}},\ }\bibfield  {title}
	{\bibinfo {title} {Boltzmann theory of engineered anisotropic
			magnetoresistance in $(\mathrm{Ga,Mn})\mathrm{As}$},\ }\href@noop {}
	{\bibfield  {journal} {\bibinfo  {journal} {Applied Physics Letters}\
		}\textbf {\bibinfo {volume} {81}},\ \bibinfo {pages} {4029} (\bibinfo {year}
		{2002})}\BibitemShut {NoStop}%
	\bibitem [{\citenamefont {Ma}\ \emph {et~al.}(2023)\citenamefont {Ma},
		\citenamefont {Huang}, \citenamefont {Wang}, \citenamefont {Liu},
		\citenamefont {Zhang}, \citenamefont {Lu},\ and\ \citenamefont
		{Xiang}}]{ma2023anisotropic}%
	\BibitemOpen
	\bibfield  {author} {\bibinfo {author} {\bibfnamefont {X.}~\bibnamefont
			{Ma}}, \bibinfo {author} {\bibfnamefont {M.}~\bibnamefont {Huang}}, \bibinfo
		{author} {\bibfnamefont {S.}~\bibnamefont {Wang}}, \bibinfo {author}
		{\bibfnamefont {P.}~\bibnamefont {Liu}}, \bibinfo {author} {\bibfnamefont
			{Y.}~\bibnamefont {Zhang}}, \bibinfo {author} {\bibfnamefont
			{Y.}~\bibnamefont {Lu}},\ and\ \bibinfo {author} {\bibfnamefont
			{B.}~\bibnamefont {Xiang}},\ }\bibfield  {title} {\bibinfo {title}
		{Anisotropic magnetoresistance and planar hall effect in layered
			room-temperature ferromagnet $\mathrm{Cr}_{1.2}\mathrm{Te}_2$},\ }\href@noop
	{} {\bibfield  {journal} {\bibinfo  {journal} {ACS Applied Electronic
				Materials}\ }\textbf {\bibinfo {volume} {5}},\ \bibinfo {pages} {2838}
		(\bibinfo {year} {2023})}\BibitemShut {NoStop}%
	\bibitem [{\citenamefont {Sun}\ \emph {et~al.}(2022)\citenamefont {Sun},
		\citenamefont {Liang}, \citenamefont {Liu}, \citenamefont {Shen},
		\citenamefont {Wu}, \citenamefont {Tian}, \citenamefont {Cao}, \citenamefont
		{Yang}, \citenamefont {Huang}, \citenamefont {Lin} \emph
		{et~al.}}]{sun2022anisotropic}%
	\BibitemOpen
	\bibfield  {author} {\bibinfo {author} {\bibfnamefont {S.}~\bibnamefont
			{Sun}}, \bibinfo {author} {\bibfnamefont {J.}~\bibnamefont {Liang}}, \bibinfo
		{author} {\bibfnamefont {R.}~\bibnamefont {Liu}}, \bibinfo {author}
		{\bibfnamefont {W.}~\bibnamefont {Shen}}, \bibinfo {author} {\bibfnamefont
			{H.}~\bibnamefont {Wu}}, \bibinfo {author} {\bibfnamefont {M.}~\bibnamefont
			{Tian}}, \bibinfo {author} {\bibfnamefont {L.}~\bibnamefont {Cao}}, \bibinfo
		{author} {\bibfnamefont {Y.}~\bibnamefont {Yang}}, \bibinfo {author}
		{\bibfnamefont {Z.}~\bibnamefont {Huang}}, \bibinfo {author} {\bibfnamefont
			{W.}~\bibnamefont {Lin}}, \emph {et~al.},\ }\bibfield  {title} {\bibinfo
		{title} {Anisotropic magnetoresistance in room temperature ferromagnetic
			single crystal $\mathrm{CrTe}$ flake},\ }\href@noop {} {\bibfield  {journal}
		{\bibinfo  {journal} {Journal of Alloys and Compounds}\ }\textbf {\bibinfo
			{volume} {890}},\ \bibinfo {pages} {161818} (\bibinfo {year}
		{2022})}\BibitemShut {NoStop}%
	\bibitem [{\citenamefont {Sun}\ \emph {et~al.}(2020)\citenamefont {Sun},
		\citenamefont {Li}, \citenamefont {Wang}, \citenamefont {Sui}, \citenamefont
		{Zhang}, \citenamefont {Wang}, \citenamefont {Liu}, \citenamefont {Li},
		\citenamefont {Feng}, \citenamefont {Zhong} \emph {et~al.}}]{sun2020room}%
	\BibitemOpen
	\bibfield  {author} {\bibinfo {author} {\bibfnamefont {X.}~\bibnamefont
			{Sun}}, \bibinfo {author} {\bibfnamefont {W.}~\bibnamefont {Li}}, \bibinfo
		{author} {\bibfnamefont {X.}~\bibnamefont {Wang}}, \bibinfo {author}
		{\bibfnamefont {Q.}~\bibnamefont {Sui}}, \bibinfo {author} {\bibfnamefont
			{T.}~\bibnamefont {Zhang}}, \bibinfo {author} {\bibfnamefont
			{Z.}~\bibnamefont {Wang}}, \bibinfo {author} {\bibfnamefont {L.}~\bibnamefont
			{Liu}}, \bibinfo {author} {\bibfnamefont {D.}~\bibnamefont {Li}}, \bibinfo
		{author} {\bibfnamefont {S.}~\bibnamefont {Feng}}, \bibinfo {author}
		{\bibfnamefont {S.}~\bibnamefont {Zhong}}, \emph {et~al.},\ }\bibfield
	{title} {\bibinfo {title} {Room temperature ferromagnetism in ultra-thin van
			der waals crystals of $\mathrm{1T-CrTe}_2$},\ }\href@noop {} {\bibfield
		{journal} {\bibinfo  {journal} {Nano Research}\ }\textbf {\bibinfo {volume}
			{13}},\ \bibinfo {pages} {3358} (\bibinfo {year} {2020})}\BibitemShut
	{NoStop}%
	\bibitem [{\citenamefont {Liu}\ \emph {et~al.}(2023)\citenamefont {Liu},
		\citenamefont {Ren}, \citenamefont {Zhang}, \citenamefont {Li}, \citenamefont
		{Dong},\ and\ \citenamefont {Guo}}]{apl_lb}%
	\BibitemOpen
	\bibfield  {author} {\bibinfo {author} {\bibfnamefont {B.}~\bibnamefont
			{Liu}}, \bibinfo {author} {\bibfnamefont {X.~X.}\ \bibnamefont {Ren}},
		\bibinfo {author} {\bibfnamefont {X.}~\bibnamefont {Zhang}}, \bibinfo
		{author} {\bibfnamefont {P.}~\bibnamefont {Li}}, \bibinfo {author}
		{\bibfnamefont {Y.}~\bibnamefont {Dong}},\ and\ \bibinfo {author}
		{\bibfnamefont {Z.-X.}\ \bibnamefont {Guo}},\ }\bibfield  {title} {\bibinfo
		{title} {{Electric field tunable multi-state tunnel magnetoresistances in 2D
				van der Waals magnetic heterojunctions}},\ }\href@noop {} {\bibfield
		{journal} {\bibinfo  {journal} {Applied Physics Letters}\ }\textbf {\bibinfo
			{volume} {122}},\ \bibinfo {pages} {152408} (\bibinfo {year}
		{2023})}\BibitemShut {NoStop}%
	\bibitem [{\citenamefont {Li}\ \emph {et~al.}(2022)\citenamefont {Li},
		\citenamefont {Zhou},\ and\ \citenamefont {Guo}}]{li2022intriguing}%
	\BibitemOpen
	\bibfield  {author} {\bibinfo {author} {\bibfnamefont {P.}~\bibnamefont
			{Li}}, \bibinfo {author} {\bibfnamefont {X.-S.}\ \bibnamefont {Zhou}},\ and\
		\bibinfo {author} {\bibfnamefont {Z.-X.}\ \bibnamefont {Guo}},\ }\bibfield
	{title} {\bibinfo {title} {Intriguing magnetoelectric effect in
			two-dimensional ferromagnetic/perovskite oxide ferroelectric
			heterostructure},\ }\href@noop {} {\bibfield  {journal} {\bibinfo  {journal}
			{npj Computational Materials}\ }\textbf {\bibinfo {volume} {8}},\ \bibinfo
		{pages} {20} (\bibinfo {year} {2022})}\BibitemShut {NoStop}%
	\bibitem [{\citenamefont {Zhang}\ \emph {et~al.}(2024)\citenamefont {Zhang},
		\citenamefont {Liu}, \citenamefont {Huang}, \citenamefont {Cao},
		\citenamefont {Zhang},\ and\ \citenamefont {Guo}}]{PhysRevB.109.205105}%
	\BibitemOpen
	\bibfield  {author} {\bibinfo {author} {\bibfnamefont {X.}~\bibnamefont
			{Zhang}}, \bibinfo {author} {\bibfnamefont {B.}~\bibnamefont {Liu}}, \bibinfo
		{author} {\bibfnamefont {J.}~\bibnamefont {Huang}}, \bibinfo {author}
		{\bibfnamefont {X.}~\bibnamefont {Cao}}, \bibinfo {author} {\bibfnamefont
			{Y.}~\bibnamefont {Zhang}},\ and\ \bibinfo {author} {\bibfnamefont {Z.-X.}\
			\bibnamefont {Guo}},\ }\bibfield  {title} {\bibinfo {title} {Nonvolatile spin
			field effect transistor based on
			$\mathrm{VSi}_{2}\mathrm{N}_{4}/\mathrm{Sc}_{2}\mathrm{CO}_{2}$ multiferroic
			heterostructure},\ }\href@noop {} {\bibfield  {journal} {\bibinfo  {journal}
			{Physical Review B}\ }\textbf {\bibinfo {volume} {109}},\ \bibinfo {pages}
		{205105} (\bibinfo {year} {2024})}\BibitemShut {NoStop}%
	\bibitem [{\citenamefont {Zhang}\ \emph {et~al.}(2021)\citenamefont {Zhang},
		\citenamefont {Yang}, \citenamefont {Wang},\ and\ \citenamefont
		{Xu}}]{PhysRevB.103.094433}%
	\BibitemOpen
	\bibfield  {author} {\bibinfo {author} {\bibfnamefont {H.}~\bibnamefont
			{Zhang}}, \bibinfo {author} {\bibfnamefont {W.}~\bibnamefont {Yang}},
		\bibinfo {author} {\bibfnamefont {Y.}~\bibnamefont {Wang}},\ and\ \bibinfo
		{author} {\bibfnamefont {X.}~\bibnamefont {Xu}},\ }\bibfield  {title}
	{\bibinfo {title} {Tunable topological states in layered magnetic materials
			of $\mathrm{Mn}\mathrm{Sb}_{2}\mathrm{Te}_{4},
			\mathrm{Mn}\mathrm{Bi}_{2}\mathrm{Se}_{4}$, and
			$\mathrm{Mn}\mathrm{Sb}_{2}\mathrm{Se}_{4}$},\ }\href@noop {} {\bibfield
		{journal} {\bibinfo  {journal} {Physical Review B}\ }\textbf {\bibinfo
			{volume} {103}},\ \bibinfo {pages} {094433} (\bibinfo {year}
		{2021})}\BibitemShut {NoStop}%
	\bibitem [{\citenamefont {Zhang}\ \emph {et~al.}(2020)\citenamefont {Zhang},
		\citenamefont {Yang}, \citenamefont {Cui}, \citenamefont {Xu},\ and\
		\citenamefont {Zhang}}]{PhysRevB.102.115413}%
	\BibitemOpen
	\bibfield  {author} {\bibinfo {author} {\bibfnamefont {H.}~\bibnamefont
			{Zhang}}, \bibinfo {author} {\bibfnamefont {W.}~\bibnamefont {Yang}},
		\bibinfo {author} {\bibfnamefont {P.}~\bibnamefont {Cui}}, \bibinfo {author}
		{\bibfnamefont {X.}~\bibnamefont {Xu}},\ and\ \bibinfo {author}
		{\bibfnamefont {Z.}~\bibnamefont {Zhang}},\ }\bibfield  {title} {\bibinfo
		{title} {Prediction of monolayered ferromagnetic ${\mathrm{crmni}}_{6}$ as an
			intrinsic high-temperature quantum anomalous hall system},\ }\href@noop {}
	{\bibfield  {journal} {\bibinfo  {journal} {Physical Review B}\ }\textbf
		{\bibinfo {volume} {102}},\ \bibinfo {pages} {115413} (\bibinfo {year}
		{2020})}\BibitemShut {NoStop}%
	\bibitem [{\citenamefont {Burch}\ \emph {et~al.}(2018)\citenamefont {Burch},
		\citenamefont {Mandrus},\ and\ \citenamefont {Park}}]{burch2018magnetism}%
	\BibitemOpen
	\bibfield  {author} {\bibinfo {author} {\bibfnamefont {K.~S.}\ \bibnamefont
			{Burch}}, \bibinfo {author} {\bibfnamefont {D.}~\bibnamefont {Mandrus}},\
		and\ \bibinfo {author} {\bibfnamefont {J.-G.}\ \bibnamefont {Park}},\
	}\bibfield  {title} {\bibinfo {title} {Magnetism in two-dimensional van der
			waals materials},\ }\href@noop {} {\bibfield  {journal} {\bibinfo  {journal}
			{Nature}\ }\textbf {\bibinfo {volume} {563}},\ \bibinfo {pages} {47}
		(\bibinfo {year} {2018})}\BibitemShut {NoStop}%
	\bibitem [{\citenamefont {Kresse}\ and\ \citenamefont
		{Joubert}(1999)}]{PhysRevB.59.1758}%
	\BibitemOpen
	\bibfield  {author} {\bibinfo {author} {\bibfnamefont {G.}~\bibnamefont
			{Kresse}}\ and\ \bibinfo {author} {\bibfnamefont {D.}~\bibnamefont
			{Joubert}},\ }\bibfield  {title} {\bibinfo {title} {From ultrasoft
			pseudopotentials to the projector augmented-wave method},\ }\href@noop {}
	{\bibfield  {journal} {\bibinfo  {journal} {Physical Review B}\ }\textbf
		{\bibinfo {volume} {59}},\ \bibinfo {pages} {1758} (\bibinfo {year}
		{1999})}\BibitemShut {NoStop}%
	\bibitem [{\citenamefont {Bl\"ochl}(1994)}]{PhysRevB.50.17953}%
	\BibitemOpen
	\bibfield  {author} {\bibinfo {author} {\bibfnamefont {P.~E.}\ \bibnamefont
			{Bl\"ochl}},\ }\bibfield  {title} {\bibinfo {title} {Projector augmented-wave
			method},\ }\href@noop {} {\bibfield  {journal} {\bibinfo  {journal} {Physical
				Review B}\ }\textbf {\bibinfo {volume} {50}},\ \bibinfo {pages} {17953}
		(\bibinfo {year} {1994})}\BibitemShut {NoStop}%
	\bibitem [{\citenamefont {Kresse}\ and\ \citenamefont
		{Furthm\"uller}(1996)}]{PhysRevB.54.11169}%
	\BibitemOpen
	\bibfield  {author} {\bibinfo {author} {\bibfnamefont {G.}~\bibnamefont
			{Kresse}}\ and\ \bibinfo {author} {\bibfnamefont {J.}~\bibnamefont
			{Furthm\"uller}},\ }\bibfield  {title} {\bibinfo {title} {Efficient iterative
			schemes for ab initio total-energy calculations using a plane-wave basis
			set},\ }\href@noop {} {\bibfield  {journal} {\bibinfo  {journal} {Physical
				Review B}\ }\textbf {\bibinfo {volume} {54}},\ \bibinfo {pages} {11169}
		(\bibinfo {year} {1996})}\BibitemShut {NoStop}%
	\bibitem [{\citenamefont {Kresse}\ and\ \citenamefont
		{Hafner}(1993)}]{PhysRevB.47.558}%
	\BibitemOpen
	\bibfield  {author} {\bibinfo {author} {\bibfnamefont {G.}~\bibnamefont
			{Kresse}}\ and\ \bibinfo {author} {\bibfnamefont {J.}~\bibnamefont
			{Hafner}},\ }\bibfield  {title} {\bibinfo {title} {Ab initio molecular
			dynamics for liquid metals},\ }\href@noop {} {\bibfield  {journal} {\bibinfo
			{journal} {Physical Review B}\ }\textbf {\bibinfo {volume} {47}},\ \bibinfo
		{pages} {558} (\bibinfo {year} {1993})}\BibitemShut {NoStop}%
	\bibitem [{\citenamefont {Wannier}(1937)}]{PhysRev.52.191}%
	\BibitemOpen
	\bibfield  {author} {\bibinfo {author} {\bibfnamefont {G.~H.}\ \bibnamefont
			{Wannier}},\ }\bibfield  {title} {\bibinfo {title} {The structure of
			electronic excitation levels in insulating crystals},\ }\href@noop {}
	{\bibfield  {journal} {\bibinfo  {journal} {Physical Review}\ }\textbf
		{\bibinfo {volume} {52}},\ \bibinfo {pages} {191} (\bibinfo {year}
		{1937})}\BibitemShut {NoStop}%
	\bibitem [{\citenamefont {Marzari}\ \emph {et~al.}(2012)\citenamefont
		{Marzari}, \citenamefont {Mostofi}, \citenamefont {Yates}, \citenamefont
		{Souza},\ and\ \citenamefont {Vanderbilt}}]{RevModPhys.84.1419}%
	\BibitemOpen
	\bibfield  {author} {\bibinfo {author} {\bibfnamefont {N.}~\bibnamefont
			{Marzari}}, \bibinfo {author} {\bibfnamefont {A.~A.}\ \bibnamefont
			{Mostofi}}, \bibinfo {author} {\bibfnamefont {J.~R.}\ \bibnamefont {Yates}},
		\bibinfo {author} {\bibfnamefont {I.}~\bibnamefont {Souza}},\ and\ \bibinfo
		{author} {\bibfnamefont {D.}~\bibnamefont {Vanderbilt}},\ }\bibfield  {title}
	{\bibinfo {title} {Maximally localized wannier functions: Theory and
			applications},\ }\href@noop {} {\bibfield  {journal} {\bibinfo  {journal}
			{Reviews of Modern Physics}\ }\textbf {\bibinfo {volume} {84}},\ \bibinfo
		{pages} {1419} (\bibinfo {year} {2012})}\BibitemShut {NoStop}%
	\bibitem [{\citenamefont {Marzari}\ and\ \citenamefont
		{Vanderbilt}(1997)}]{PhysRevB.56.12847}%
	\BibitemOpen
	\bibfield  {author} {\bibinfo {author} {\bibfnamefont {N.}~\bibnamefont
			{Marzari}}\ and\ \bibinfo {author} {\bibfnamefont {D.}~\bibnamefont
			{Vanderbilt}},\ }\bibfield  {title} {\bibinfo {title} {Maximally localized
			generalized wannier functions for composite energy bands},\ }\href@noop {}
	{\bibfield  {journal} {\bibinfo  {journal} {Physical Review B}\ }\textbf
		{\bibinfo {volume} {56}},\ \bibinfo {pages} {12847} (\bibinfo {year}
		{1997})}\BibitemShut {NoStop}%
	\bibitem [{\citenamefont {Mostofi}\ \emph {et~al.}(2008)\citenamefont
		{Mostofi}, \citenamefont {Yates}, \citenamefont {Lee}, \citenamefont {Souza},
		\citenamefont {Vanderbilt},\ and\ \citenamefont {Marzari}}]{MOSTOFI2008685}%
	\BibitemOpen
	\bibfield  {author} {\bibinfo {author} {\bibfnamefont {A.~A.}\ \bibnamefont
			{Mostofi}}, \bibinfo {author} {\bibfnamefont {J.~R.}\ \bibnamefont {Yates}},
		\bibinfo {author} {\bibfnamefont {Y.-S.}\ \bibnamefont {Lee}}, \bibinfo
		{author} {\bibfnamefont {I.}~\bibnamefont {Souza}}, \bibinfo {author}
		{\bibfnamefont {D.}~\bibnamefont {Vanderbilt}},\ and\ \bibinfo {author}
		{\bibfnamefont {N.}~\bibnamefont {Marzari}},\ }\bibfield  {title} {\bibinfo
		{title} {wannier90: A tool for obtaining maximally-localised wannier
			functions},\ }\href
	{https://doi.org/https://doi.org/10.1016/j.cpc.2007.11.016} {\bibfield
		{journal} {\bibinfo  {journal} {Computer Physics Communications}\ }\textbf
		{\bibinfo {volume} {178}},\ \bibinfo {pages} {685} (\bibinfo {year}
		{2008})}\BibitemShut {NoStop}%
	\bibitem [{\citenamefont {Pizzi}\ \emph {et~al.}(2020)\citenamefont {Pizzi},
		\citenamefont {Vitale}, \citenamefont {Arita}, \citenamefont {Blügel},
		\citenamefont {Freimuth}, \citenamefont {Géranton}, \citenamefont
		{Gibertini}, \citenamefont {Gresch}, \citenamefont {Johnson}, \citenamefont
		{Koretsune}, \citenamefont {Ibañez-Azpiroz}, \citenamefont {Lee},
		\citenamefont {Lihm}, \citenamefont {Marchand}, \citenamefont {Marrazzo},
		\citenamefont {Mokrousov}, \citenamefont {Mustafa}, \citenamefont {Nohara},
		\citenamefont {Nomura}, \citenamefont {Paulatto}, \citenamefont {Poncé},
		\citenamefont {Ponweiser}, \citenamefont {Qiao}, \citenamefont {Thöle},
		\citenamefont {Tsirkin}, \citenamefont {Wierzbowska}, \citenamefont
		{Marzari}, \citenamefont {Vanderbilt}, \citenamefont {Souza}, \citenamefont
		{Mostofi},\ and\ \citenamefont {Yates}}]{Pizzi_2020}%
	\BibitemOpen
	\bibfield  {author} {\bibinfo {author} {\bibfnamefont {G.}~\bibnamefont
			{Pizzi}}, \bibinfo {author} {\bibfnamefont {V.}~\bibnamefont {Vitale}},
		\bibinfo {author} {\bibfnamefont {R.}~\bibnamefont {Arita}}, \bibinfo
		{author} {\bibfnamefont {S.}~\bibnamefont {Blügel}}, \bibinfo {author}
		{\bibfnamefont {F.}~\bibnamefont {Freimuth}}, \bibinfo {author}
		{\bibfnamefont {G.}~\bibnamefont {Géranton}}, \bibinfo {author}
		{\bibfnamefont {M.}~\bibnamefont {Gibertini}}, \bibinfo {author}
		{\bibfnamefont {D.}~\bibnamefont {Gresch}}, \bibinfo {author} {\bibfnamefont
			{C.}~\bibnamefont {Johnson}}, \bibinfo {author} {\bibfnamefont
			{T.}~\bibnamefont {Koretsune}}, \bibinfo {author} {\bibfnamefont
			{J.}~\bibnamefont {Ibañez-Azpiroz}}, \bibinfo {author} {\bibfnamefont
			{H.}~\bibnamefont {Lee}}, \bibinfo {author} {\bibfnamefont {J.-M.}\
			\bibnamefont {Lihm}}, \bibinfo {author} {\bibfnamefont {D.}~\bibnamefont
			{Marchand}}, \bibinfo {author} {\bibfnamefont {A.}~\bibnamefont {Marrazzo}},
		\bibinfo {author} {\bibfnamefont {Y.}~\bibnamefont {Mokrousov}}, \bibinfo
		{author} {\bibfnamefont {J.~I.}\ \bibnamefont {Mustafa}}, \bibinfo {author}
		{\bibfnamefont {Y.}~\bibnamefont {Nohara}}, \bibinfo {author} {\bibfnamefont
			{Y.}~\bibnamefont {Nomura}}, \bibinfo {author} {\bibfnamefont
			{L.}~\bibnamefont {Paulatto}}, \bibinfo {author} {\bibfnamefont
			{S.}~\bibnamefont {Poncé}}, \bibinfo {author} {\bibfnamefont
			{T.}~\bibnamefont {Ponweiser}}, \bibinfo {author} {\bibfnamefont
			{J.}~\bibnamefont {Qiao}}, \bibinfo {author} {\bibfnamefont {F.}~\bibnamefont
			{Thöle}}, \bibinfo {author} {\bibfnamefont {S.~S.}\ \bibnamefont {Tsirkin}},
		\bibinfo {author} {\bibfnamefont {M.}~\bibnamefont {Wierzbowska}}, \bibinfo
		{author} {\bibfnamefont {N.}~\bibnamefont {Marzari}}, \bibinfo {author}
		{\bibfnamefont {D.}~\bibnamefont {Vanderbilt}}, \bibinfo {author}
		{\bibfnamefont {I.}~\bibnamefont {Souza}}, \bibinfo {author} {\bibfnamefont
			{A.~A.}\ \bibnamefont {Mostofi}},\ and\ \bibinfo {author} {\bibfnamefont
			{J.~R.}\ \bibnamefont {Yates}},\ }\bibfield  {title} {\bibinfo {title}
		{Wannier90 as a community code: new features and applications},\ }\href
	{https://doi.org/10.1088/1361-648X/ab51ff} {\bibfield  {journal} {\bibinfo
			{journal} {Journal of Physics: Condensed Matter}\ }\textbf {\bibinfo {volume}
			{32}},\ \bibinfo {pages} {165902} (\bibinfo {year} {2020})}\BibitemShut
	{NoStop}%
	\bibitem [{\citenamefont {Ziman}(1972)}]{Ziman_1972}%
	\BibitemOpen
	\bibfield  {author} {\bibinfo {author} {\bibfnamefont {J.~M.}\ \bibnamefont
			{Ziman}},\ }\href@noop {} {\emph {\bibinfo {title} {Principles of the Theory
				of Solids}}},\ \bibinfo {edition} {2nd}\ ed.\ (\bibinfo  {publisher}
	{Cambridge University Press},\ \bibinfo {year} {1972})\BibitemShut {NoStop}%
	\bibitem [{\citenamefont {Grosso}\ and\ \citenamefont
		{Parravicini}(2013)}]{grosso2013solid}%
	\BibitemOpen
	\bibfield  {author} {\bibinfo {author} {\bibfnamefont {G.}~\bibnamefont
			{Grosso}}\ and\ \bibinfo {author} {\bibfnamefont {G.~P.}\ \bibnamefont
			{Parravicini}},\ }\href@noop {} {\emph {\bibinfo {title} {Solid state
				physics}}}\ (\bibinfo  {publisher} {Academic press},\ \bibinfo {year}
	{2013})\BibitemShut {NoStop}%
	\bibitem [{\citenamefont {Pizzi}\ \emph {et~al.}(2014)\citenamefont {Pizzi},
		\citenamefont {Volja}, \citenamefont {Kozinsky}, \citenamefont {Fornari},\
		and\ \citenamefont {Marzari}}]{PIZZI2014422}%
	\BibitemOpen
	\bibfield  {author} {\bibinfo {author} {\bibfnamefont {G.}~\bibnamefont
			{Pizzi}}, \bibinfo {author} {\bibfnamefont {D.}~\bibnamefont {Volja}},
		\bibinfo {author} {\bibfnamefont {B.}~\bibnamefont {Kozinsky}}, \bibinfo
		{author} {\bibfnamefont {M.}~\bibnamefont {Fornari}},\ and\ \bibinfo {author}
		{\bibfnamefont {N.}~\bibnamefont {Marzari}},\ }\bibfield  {title} {\bibinfo
		{title} {Boltzwann: A code for the evaluation of thermoelectric and
			electronic transport properties with a maximally-localized wannier functions
			basis},\ }\href {https://doi.org/https://doi.org/10.1016/j.cpc.2013.09.015}
	{\bibfield  {journal} {\bibinfo  {journal} {Computer Physics Communications}\
		}\textbf {\bibinfo {volume} {185}},\ \bibinfo {pages} {422} (\bibinfo {year}
		{2014})}\BibitemShut {NoStop}%
	\bibitem [{sup()}]{supplementary}%
	\BibitemOpen
	\bibfield  {title} {\bibinfo {title} {{See Supplemental Material at
				http://link.aps.org/xxx for the details of Computational Method; Atomic
				structures of monolayer $\text{CrTe}_2$; Band structures and berry curvature
				of monolayer $\text{Fe}_3\text{GeTe}_3$}},\ }\href@noop {} {\ }\BibitemShut
	{NoStop}%
	\bibitem [{\citenamefont {Deng}\ \emph {et~al.}(2018)\citenamefont {Deng},
		\citenamefont {Yu}, \citenamefont {Song}, \citenamefont {Zhang},
		\citenamefont {Wang}, \citenamefont {Sun}, \citenamefont {Yi}, \citenamefont
		{Wu}, \citenamefont {Wu}, \citenamefont {Zhu} \emph {et~al.}}]{deng2018gate}%
	\BibitemOpen
	\bibfield  {author} {\bibinfo {author} {\bibfnamefont {Y.}~\bibnamefont
			{Deng}}, \bibinfo {author} {\bibfnamefont {Y.}~\bibnamefont {Yu}}, \bibinfo
		{author} {\bibfnamefont {Y.}~\bibnamefont {Song}}, \bibinfo {author}
		{\bibfnamefont {J.}~\bibnamefont {Zhang}}, \bibinfo {author} {\bibfnamefont
			{N.~Z.}\ \bibnamefont {Wang}}, \bibinfo {author} {\bibfnamefont
			{Z.}~\bibnamefont {Sun}}, \bibinfo {author} {\bibfnamefont {Y.}~\bibnamefont
			{Yi}}, \bibinfo {author} {\bibfnamefont {Y.~Z.}\ \bibnamefont {Wu}}, \bibinfo
		{author} {\bibfnamefont {S.}~\bibnamefont {Wu}}, \bibinfo {author}
		{\bibfnamefont {J.}~\bibnamefont {Zhu}}, \emph {et~al.},\ }\bibfield  {title}
	{\bibinfo {title} {Gate-tunable room-temperature ferromagnetism in
			two-dimensional $\text{Fe}_3\text{GeTe}_2$},\ }\href@noop {} {\bibfield
		{journal} {\bibinfo  {journal} {Nature}\ }\textbf {\bibinfo {volume} {563}},\
		\bibinfo {pages} {94} (\bibinfo {year} {2018})}\BibitemShut {NoStop}%
	\bibitem [{\citenamefont {Fei}\ \emph {et~al.}(2018)\citenamefont {Fei},
		\citenamefont {Huang}, \citenamefont {Malinowski}, \citenamefont {Wang},
		\citenamefont {Song}, \citenamefont {Sanchez}, \citenamefont {Yao},
		\citenamefont {Xiao}, \citenamefont {Zhu}, \citenamefont {May} \emph
		{et~al.}}]{fei2018two}%
	\BibitemOpen
	\bibfield  {author} {\bibinfo {author} {\bibfnamefont {Z.}~\bibnamefont
			{Fei}}, \bibinfo {author} {\bibfnamefont {B.}~\bibnamefont {Huang}}, \bibinfo
		{author} {\bibfnamefont {P.}~\bibnamefont {Malinowski}}, \bibinfo {author}
		{\bibfnamefont {W.}~\bibnamefont {Wang}}, \bibinfo {author} {\bibfnamefont
			{T.}~\bibnamefont {Song}}, \bibinfo {author} {\bibfnamefont {J.}~\bibnamefont
			{Sanchez}}, \bibinfo {author} {\bibfnamefont {W.}~\bibnamefont {Yao}},
		\bibinfo {author} {\bibfnamefont {D.}~\bibnamefont {Xiao}}, \bibinfo {author}
		{\bibfnamefont {X.}~\bibnamefont {Zhu}}, \bibinfo {author} {\bibfnamefont
			{A.~F.}\ \bibnamefont {May}}, \emph {et~al.},\ }\bibfield  {title} {\bibinfo
		{title} {Two-dimensional itinerant ferromagnetism in atomically thin
			$\text{Fe}_3\text{GeTe}_2$},\ }\href@noop {} {\bibfield  {journal} {\bibinfo
			{journal} {Nature materials}\ }\textbf {\bibinfo {volume} {17}},\ \bibinfo
		{pages} {778} (\bibinfo {year} {2018})}\BibitemShut {NoStop}%
	\bibitem [{\citenamefont {Zhuang}\ \emph {et~al.}(2016)\citenamefont {Zhuang},
		\citenamefont {Kent},\ and\ \citenamefont {Hennig}}]{PhysRevB.93.134407}%
	\BibitemOpen
	\bibfield  {author} {\bibinfo {author} {\bibfnamefont {H.~L.}\ \bibnamefont
			{Zhuang}}, \bibinfo {author} {\bibfnamefont {P.~R.~C.}\ \bibnamefont
			{Kent}},\ and\ \bibinfo {author} {\bibfnamefont {R.~G.}\ \bibnamefont
			{Hennig}},\ }\bibfield  {title} {\bibinfo {title} {Strong anisotropy and
			magnetostriction in the two-dimensional stoner ferromagnet
			$\text{Fe}_3\text{GeTe}_2$},\ }\href@noop {} {\bibfield  {journal} {\bibinfo
			{journal} {Physical Review B}\ }\textbf {\bibinfo {volume} {93}},\ \bibinfo
		{pages} {134407} (\bibinfo {year} {2016})}\BibitemShut {NoStop}%
	\bibitem [{\citenamefont {Deiseroth}\ \emph {et~al.}(2006)\citenamefont
		{Deiseroth}, \citenamefont {Aleksandrov}, \citenamefont {Reiner},
		\citenamefont {Kienle},\ and\ \citenamefont
		{Kremer}}]{deiseroth2006fe3gete2}%
	\BibitemOpen
	\bibfield  {author} {\bibinfo {author} {\bibfnamefont {H.-J.}\ \bibnamefont
			{Deiseroth}}, \bibinfo {author} {\bibfnamefont {K.}~\bibnamefont
			{Aleksandrov}}, \bibinfo {author} {\bibfnamefont {C.}~\bibnamefont {Reiner}},
		\bibinfo {author} {\bibfnamefont {L.}~\bibnamefont {Kienle}},\ and\ \bibinfo
		{author} {\bibfnamefont {R.~K.}\ \bibnamefont {Kremer}},\ }\href@noop {}
	{\bibinfo {title} {$\text{Fe}_3\text{GeTe}_2$ and
			$\text{Ni}_3\text{GeTe}_2$--two new layered transition-metal compounds:
			crystal structures, hrtem investigations, and magnetic and electrical
			properties}} (\bibinfo {year} {2006})\BibitemShut {NoStop}%
	\bibitem [{\citenamefont {Dong}\ \emph {et~al.}(2023)\citenamefont {Dong},
		\citenamefont {Guo},\ and\ \citenamefont {Wang}}]{PhysRevB.108.L020401}%
	\BibitemOpen
	\bibfield  {author} {\bibinfo {author} {\bibfnamefont {M.~Q.}\ \bibnamefont
			{Dong}}, \bibinfo {author} {\bibfnamefont {Z.-X.}\ \bibnamefont {Guo}},\ and\
		\bibinfo {author} {\bibfnamefont {X.~R.}\ \bibnamefont {Wang}},\ }\bibfield
	{title} {\bibinfo {title} {Anisotropic magnetoresistance due to
			magnetization-dependent spin-orbit interactions},\ }\href@noop {} {\bibfield
		{journal} {\bibinfo  {journal} {Physical Review B}\ }\textbf {\bibinfo
			{volume} {108}},\ \bibinfo {pages} {L020401} (\bibinfo {year}
		{2023})}\BibitemShut {NoStop}%
	\bibitem [{\citenamefont {Hou}\ \emph {et~al.}(2024)\citenamefont {Hou},
		\citenamefont {Dong},\ and\ \citenamefont
		{Guo}}]{hou2024giantanisotropicmagnetoresistancemagnetic}%
	\BibitemOpen
	\bibfield  {author} {\bibinfo {author} {\bibfnamefont {W.~S.}\ \bibnamefont
			{Hou}}, \bibinfo {author} {\bibfnamefont {M.~Q.}\ \bibnamefont {Dong}},\ and\
		\bibinfo {author} {\bibfnamefont {Z.-X.}\ \bibnamefont {Guo}},\ }\href
	{https://arxiv.org/abs/2407.10438} {\bibinfo {title} {Giant anisotropic
			magnetoresistance in magnetic monolayers $\text{CrPX}_3 \text{(X = S, Se,
				Te)}$ due to symmetry breaking between the in-plane and out-of-plane
			crystallographic axes}} (\bibinfo {year} {2024}),\ \Eprint
	{https://arxiv.org/abs/2407.10438} {arXiv:2407.10438 [cond-mat.mes-hall]}
	\BibitemShut {NoStop}%
	\bibitem [{\citenamefont {Gonzalez~Betancourt}\ \emph
		{et~al.}(2024)\citenamefont {Gonzalez~Betancourt}, \citenamefont
		{Zub\'a\v{c}}, \citenamefont {Geishendorf}, \citenamefont {Ritzinger},
		\citenamefont {R$\ensuremath\ddot{\text{u}}${\v{z}}i\v{c}kov\'a},
		\citenamefont {Kotte}, \citenamefont {\v{Z}elezn\`y},\ and\ \citenamefont
		{Olejn\'ik}}]{gonzalez2024anisotropic}%
	\BibitemOpen
	\bibfield  {author} {\bibinfo {author} {\bibfnamefont {R.~D.}\ \bibnamefont
			{Gonzalez~Betancourt}}, \bibinfo {author} {\bibfnamefont {J.}~\bibnamefont
			{Zub\'a\v{c}}}, \bibinfo {author} {\bibfnamefont {K.}~\bibnamefont
			{Geishendorf}}, \bibinfo {author} {\bibfnamefont {P.}~\bibnamefont
			{Ritzinger}}, \bibinfo {author} {\bibfnamefont {B.}~\bibnamefont
			{R$\ensuremath\ddot{\text{u}}${\v{z}}i\v{c}kov\'a}}, \bibinfo {author}
		{\bibfnamefont {T.}~\bibnamefont {Kotte}}, \bibinfo {author} {\bibfnamefont
			{J.}~\bibnamefont {\v{Z}elezn\`y}},\ and\ \bibinfo {author} {\bibfnamefont
			{K.}~\bibnamefont {Olejn\'ik}},\ }\bibfield  {title} {\bibinfo {title}
		{Anisotropic magnetoresistance in altermagnetic {MnTe}},\ }\href@noop {}
	{\bibfield  {journal} {\bibinfo  {journal} {npj Spintronics}\ }\textbf
		{\bibinfo {volume} {2}},\ \bibinfo {pages} {45} (\bibinfo {year}
		{2024})}\BibitemShut {NoStop}%
	\bibitem [{\citenamefont {Zhang}\ \emph {et~al.}(2019)\citenamefont {Zhang},
		\citenamefont {Wu}, \citenamefont {Liu},\ and\ \citenamefont
		{Yazyev}}]{PhysRevB.99.035142}%
	\BibitemOpen
	\bibfield  {author} {\bibinfo {author} {\bibfnamefont {S.}~\bibnamefont
			{Zhang}}, \bibinfo {author} {\bibfnamefont {Q.}~\bibnamefont {Wu}}, \bibinfo
		{author} {\bibfnamefont {Y.}~\bibnamefont {Liu}},\ and\ \bibinfo {author}
		{\bibfnamefont {O.~V.}\ \bibnamefont {Yazyev}},\ }\bibfield  {title}
	{\bibinfo {title} {Magnetoresistance from fermi surface topology},\
	}\href@noop {} {\bibfield  {journal} {\bibinfo  {journal} {Physical Review
				B}\ }\textbf {\bibinfo {volume} {99}},\ \bibinfo {pages} {035142} (\bibinfo
		{year} {2019})}\BibitemShut {NoStop}%
	\bibitem [{\citenamefont {Li}\ \emph {et~al.}(2017)\citenamefont {Li},
		\citenamefont {Li}, \citenamefont {Zhao}, \citenamefont {Chen}, \citenamefont
		{Chen}, \citenamefont {Guo}, \citenamefont {Feng}, \citenamefont {Gong},\
		and\ \citenamefont {MacDonald}}]{li2017topological}%
	\BibitemOpen
	\bibfield  {author} {\bibinfo {author} {\bibfnamefont {P.}~\bibnamefont
			{Li}}, \bibinfo {author} {\bibfnamefont {X.}~\bibnamefont {Li}}, \bibinfo
		{author} {\bibfnamefont {W.}~\bibnamefont {Zhao}}, \bibinfo {author}
		{\bibfnamefont {H.}~\bibnamefont {Chen}}, \bibinfo {author} {\bibfnamefont
			{M.-X.}\ \bibnamefont {Chen}}, \bibinfo {author} {\bibfnamefont {Z.-X.}\
			\bibnamefont {Guo}}, \bibinfo {author} {\bibfnamefont {J.}~\bibnamefont
			{Feng}}, \bibinfo {author} {\bibfnamefont {X.-G.}\ \bibnamefont {Gong}},\
		and\ \bibinfo {author} {\bibfnamefont {A.~H.}\ \bibnamefont {MacDonald}},\
	}\bibfield  {title} {\bibinfo {title} {Topological dirac states beyond
			$\pi$-orbitals for silicene on sic (0001) surface},\ }\href@noop {}
	{\bibfield  {journal} {\bibinfo  {journal} {Nano Letters}\ }\textbf {\bibinfo
			{volume} {17}},\ \bibinfo {pages} {6195} (\bibinfo {year}
		{2017})}\BibitemShut {NoStop}%
	\bibitem [{\citenamefont {Sundaram}\ and\ \citenamefont
		{Niu}(1999)}]{PhysRevB.59.14915}%
	\BibitemOpen
	\bibfield  {author} {\bibinfo {author} {\bibfnamefont {G.}~\bibnamefont
			{Sundaram}}\ and\ \bibinfo {author} {\bibfnamefont {Q.}~\bibnamefont {Niu}},\
	}\bibfield  {title} {\bibinfo {title} {Wave-packet dynamics in slowly
			perturbed crystals: Gradient corrections and berry-phase effects},\
	}\href@noop {} {\bibfield  {journal} {\bibinfo  {journal} {Physical Review
				B}\ }\textbf {\bibinfo {volume} {59}},\ \bibinfo {pages} {14915} (\bibinfo
		{year} {1999})}\BibitemShut {NoStop}%
	\bibitem [{\citenamefont {Chang}\ and\ \citenamefont
		{Niu}(1996)}]{PhysRevB.53.7010}%
	\BibitemOpen
	\bibfield  {author} {\bibinfo {author} {\bibfnamefont {M.-C.}\ \bibnamefont
			{Chang}}\ and\ \bibinfo {author} {\bibfnamefont {Q.}~\bibnamefont {Niu}},\
	}\bibfield  {title} {\bibinfo {title} {Berry phase, hyperorbits, and the
			hofstadter spectrum: Semiclassical dynamics in magnetic bloch bands},\
	}\href@noop {} {\bibfield  {journal} {\bibinfo  {journal} {Physical Review
				B}\ }\textbf {\bibinfo {volume} {53}},\ \bibinfo {pages} {7010} (\bibinfo
		{year} {1996})}\BibitemShut {NoStop}%
	\bibitem [{\citenamefont {Yao}\ \emph {et~al.}(2004)\citenamefont {Yao},
		\citenamefont {Kleinman}, \citenamefont {MacDonald}, \citenamefont {Sinova},
		\citenamefont {Jungwirth}, \citenamefont {Wang}, \citenamefont {Wang},\ and\
		\citenamefont {Niu}}]{PhysRevLett.92.037204}%
	\BibitemOpen
	\bibfield  {author} {\bibinfo {author} {\bibfnamefont {Y.}~\bibnamefont
			{Yao}}, \bibinfo {author} {\bibfnamefont {L.}~\bibnamefont {Kleinman}},
		\bibinfo {author} {\bibfnamefont {A.~H.}\ \bibnamefont {MacDonald}}, \bibinfo
		{author} {\bibfnamefont {J.}~\bibnamefont {Sinova}}, \bibinfo {author}
		{\bibfnamefont {T.}~\bibnamefont {Jungwirth}}, \bibinfo {author}
		{\bibfnamefont {D.-s.}\ \bibnamefont {Wang}}, \bibinfo {author}
		{\bibfnamefont {E.}~\bibnamefont {Wang}},\ and\ \bibinfo {author}
		{\bibfnamefont {Q.}~\bibnamefont {Niu}},\ }\bibfield  {title} {\bibinfo
		{title} {First principles calculation of anomalous hall conductivity in
			ferromagnetic bcc $\text{Fe}$},\ }\href@noop {} {\bibfield  {journal}
		{\bibinfo  {journal} {Physical Review Letters}\ }\textbf {\bibinfo {volume}
			{92}},\ \bibinfo {pages} {037204} (\bibinfo {year} {2004})}\BibitemShut
	{NoStop}%
	\bibitem [{\citenamefont {Gerlach}\ and\ \citenamefont
		{Stern}(1989)}]{gerlach1989experimentelle}%
	\BibitemOpen
	\bibfield  {author} {\bibinfo {author} {\bibfnamefont {W.}~\bibnamefont
			{Gerlach}}\ and\ \bibinfo {author} {\bibfnamefont {O.}~\bibnamefont
			{Stern}},\ }\bibfield  {title} {\bibinfo {title} {Der experimentelle nachweis
			der richtungsquantelung im magnetfeld},\ }\href@noop {} {\bibfield  {journal}
		{\bibinfo  {journal} {Walther Gerlach (1889--1979) Eine Auswahl aus seinen
				Schriften und Briefen}\ ,\ \bibinfo {pages} {26}} (\bibinfo {year}
		{1989})}\BibitemShut {NoStop}%
	\bibitem [{\citenamefont {Sinova}\ \emph {et~al.}(2015)\citenamefont {Sinova},
		\citenamefont {Valenzuela}, \citenamefont {Wunderlich}, \citenamefont
		{Back},\ and\ \citenamefont {Jungwirth}}]{RevModPhys.87.1213}%
	\BibitemOpen
	\bibfield  {author} {\bibinfo {author} {\bibfnamefont {J.}~\bibnamefont
			{Sinova}}, \bibinfo {author} {\bibfnamefont {S.~O.}\ \bibnamefont
			{Valenzuela}}, \bibinfo {author} {\bibfnamefont {J.}~\bibnamefont
			{Wunderlich}}, \bibinfo {author} {\bibfnamefont {C.~H.}\ \bibnamefont
			{Back}},\ and\ \bibinfo {author} {\bibfnamefont {T.}~\bibnamefont
			{Jungwirth}},\ }\bibfield  {title} {\bibinfo {title} {Spin hall effects},\
	}\href@noop {} {\bibfield  {journal} {\bibinfo  {journal} {Reviews of Modern
				Physics}\ }\textbf {\bibinfo {volume} {87}},\ \bibinfo {pages} {1213}
		(\bibinfo {year} {2015})}\BibitemShut {NoStop}%
\end{thebibliography}

%

\end{document}